\documentclass[12pt,preprint]{aastex}

\slugcomment{Accepted for publication in The Astrophysical Journal}

\shorttitle{$FUSE$ Abundances in I~Zw~18 ISM}
\shortauthors{Aloisi et al.}

\begin{document}


\title{Abundances in the Neutral Interstellar Medium of I~Zw~18 \\ from $FUSE$ Observations}


\author{A. Aloisi\altaffilmark{1}, S. Savaglio\altaffilmark{1,2}, 
T.M. Heckman\altaffilmark{1}, C.G. Hoopes\altaffilmark{1}, C. 
Leitherer\altaffilmark{3}, \& K.R. Sembach\altaffilmark{3}}


\altaffiltext{1}{Department of Physics \& Astronomy, Johns Hopkins University,
Baltimore, MD 21218}
\altaffiltext{2}{On leave of absence from INAF-Osservatorio Astronomico di Roma, Italy}
\altaffiltext{3}{Space Telescope Science Institute, Baltimore, MD 21218}



\begin{abstract}

We report on new $FUSE$ far-UV spectroscopy of the most metal-poor blue compact dwarf galaxy I~Zw~18. The 
new data represent an improvement over previous $FUSE$ spectra by a factor of 1.7 in the signal-to-noise. 
Together with a larger spectral coverage ($\lambda\lambda$ = 917$-$1188 \AA), this allows us to characterize 
absorption lines in the interstellar medium with unprecedented accuracy. The kinematics averaged over the 
large sampled region shows no clear evidence of gas inflows or outflows. The H~{\sc i} absorption is 
interstellar with a column density of $2.2^{+\,0.6}_{-\,0.5}\,\times\,10^{21}$ cm$^{-2}$. A conservative 
$3 \sigma$ upper limit of 5.25\,$\times$\,10$^{14}$ cm$^{-2}$ is derived for the column density of diffuse 
H$_2$. From a simultaneous fitting of metal absorption lines in the interstellar medium, we infer the 
following abundances: [Fe/H] = $-$\,1.76 $\pm$ 0.12, [O/H] = $-$\,2.06 $\pm$ 0.28, [Si/H] = $-$\,2.09 $\pm$ 
0.12, [Ar/H] = $-$\,2.27 $\pm$ 0.13, and [N/H] = $-$\,2.88 $\pm$ 0.11. This is in general several times 
lower than in the H~{\sc ii} regions. The only exception is iron, whose abundance is the same. The abundance 
pattern of the interstellar medium suggests ancient star-formation activity with an age of at least a Gyr 
that enriched the H~{\sc i} phase. Around 470 SNe~Ia are required to produce the iron content. A more recent 
episode that started 10 to several 100 Myr ago is responsible for the additional enrichment of $\alpha$-elements 
and nitrogen in the H~{\sc ii} regions.
 
\end{abstract}




\section{Introduction}

Within the well-accepted framework of hierarchical galaxy formation, dwarf ($M$ $\lesssim$ 10$^9$ 
M$_{\odot}$) galaxies are the first systems to collapse and start forming stars in the early Universe, 
supplying the building blocks from which more massive galaxies form later through merging and accretion 
(White \& Rees 1978; Dekel \& Silk 1986; Ikeuchi \& Norman 1987). Many dwarfs, however, survive and end 
up intact in the local Universe, where they still represent the most common kind of galaxy observed at 
the present time (Marzke \& da Costa 1997).

As remnants of the galaxy-formation process, present-day dwarfs may have been sites of the earliest star 
formation activity in the Universe. However, this hypothesis is challenged by the physical properties 
of blue compact dwarf (BCD) galaxies. BCDs have very blue colors typical of young stellar systems. They 
are currently experiencing intense star-formation (rates of 0.01$-$10 M$_{\odot}$ yr$^{-1}$; Thuan 1991), 
but still contain large H~{\sc i} reservoirs (a few times 10$^8$ M$_{\odot}$; Thuan \& Martin 1981). Their 
properties are consistent with relatively short duration ($<<$ a Hubble time) for the ongoing burst. 
They are also characterized by low metal abundances (between about Z$_{\odot}$/2 and Z$_{\odot}$/50 as 
inferred by H~{\sc ii} region observations; Thuan et al. 1994), indicative of a chemically unevolved 
interstellar medium (ISM), and some of them contain much less heavy-element enrichment than the majority 
of the observed high-redshift galaxies. The most metal-poor ($Z$ $\lesssim$ 1/20 Z$_{\odot}$) BCDs have 
been pointed out as good candidate ``primeval'' galaxies in the nearby Universe, with ages less than 100 
Myr as inferred from their chemical properties (Izotov \& Thuan 1999). If indeed some BCDs turn out to be 
young galaxies, their existence would support the view that star formation in low-mass systems has been 
inhibited till the present epoch (e.g., Babul \& Rees 1992). Unveiling the evolution of extreme BCDs is 
thus of primary importance for understanding galaxy formation and evolution. 

The evolution of a galaxy is driven by the evolution of its primary constituents, stars and gas. 
Stars have a key role in consuming and returning gas, producing luminous and mechanical energy, 
and synthesizing nearly all elements heavier than He in the Universe. Stars chemically influence 
the surrounding ISM depending on their mass: $\alpha$-elements (e.g., O) are mostly released by 
SNe~II from massive stars on short time scales ($\lesssim$ 50 Myr); C and N are mainly produced 
by intermediate-mass stars on longer time scales ($\gtrsim$ 300 Myr); Fe is principally supplied 
by SNe~Ia on time scales $\gtrsim$ 1 Gyr. This can lead to unusual non-solar abundance ratios 
of the ISM when coupled with the extreme star-formation histories of BCDs. A high level of star 
formation can pollute the gas with $\alpha$-elements and result in an overall $\alpha$/Fe 
enhancement. On the other hand, the contribution from SNe~Ia during a long quiescent period can 
significantly increasing the Fe content, thus considerably decreasing the $\alpha$/Fe ratio. The 
situation can also be complicated by the presence of strong galactic winds triggered by SNe~II that, 
coupled with the shallow gravitational potential of low-mass systems, can differentially eject metals 
into the intergalactic medium (e.g. De Young \& Heckman 1994; Marlowe et al. 1995). 

The chemical properties of the ISM are, therefore, a fossil record of the star-formation history of 
a galaxy. However, different spectroscopic abundance estimators sample different elements, as well 
as different phases of the ISM and different look-back times. Measured abundances in the gaseous 
component also depend on how well, and on what time scale, the ISM is mixed, and on what time scale 
freshly produced metals cool and become visible. The chemical homogeneity of the H~{\sc ii} regions  
(e.g., Kobulnicky \& Skillman 1996, 1997) suggests that mixing is quick and efficient in dwarf systems.
Emission lines from warm ($T$ $\simeq$ 10$^4$ K) ionized gas have long been the primary chemical 
diagnostic in the ISM of star-forming galaxies. This technique is highly refined and well constraining, 
especially in the determination of O abundances (e.g., Skillman 1998). H~{\sc ii} regions probe the 
state of the gas at the birth of massive stars (a few Myr ago). However, they may already be self-polluted 
with very recently produced metals (Kunth \& Sargent 1986). X-ray observations offer abundance indicators  
for the hot ($T$ $\simeq$ 10$^6$$-$10$^8$ K) phase of the ISM (Persic et al. 1998; Martin, Kobulnicky, 
\& Heckman 2002), probing the metals released during the ongoing burst that are still trapped within 
the galactic wind and are not yet mixed with the photoionized gas. UV absorption lines from neutral 
gas supply another way to infer the metallicity of the ISM. These lines sample lower temperature regions 
($T$ $\lesssim$ 10$^4$ K) which are less likely to be involved in the star-formation, thus avoiding more 
the problem of self-pollution.

Metal abundances of BCDs have been widely investigated in the past through detailed studies of H~{\sc ii} 
regions (e.g., Izotov \& Thuan 1999). However, chemical properties of other ISM phases still remain missing 
pieces of a complicated puzzle. The H~{\sc i} content in BCDs is only $\sim$ 10\% of the dynamical mass 
(which is dominated by dark matter), but amounts to $\sim$ 90\%$-$95\% of the total baryonic mass (e.g., 
Kniazev et al. 2000). As the dominant component of the baryonic matter, neutral gas could hide the bulk of 
metals. Of necessity, we must correctly address this issue in order to understand the chemical (and physical) 
evolution of BCDs. Recently, the launch of the {\it Far Ultraviolet Spectroscopic Explorer} ($FUSE$; Moos et 
al. 2000) has provided access to the rich system of far-UV absorption lines. The strong blue continua of the 
spectra, combined with the high sensitivity and resolution of $FUSE$ in the 900$-$1200 \AA~ wavelength range, 
allow us for the first time a characterization of the metal content in the neutral ISM of BCDs. Prior to 
$FUSE$, a similar analysis of the rest-frame UV spectra has been possible only for the high-redshift 
absorbing systems along the lines of sight to quasars (e.g., Lu et al. 1996) or for the gravitationally 
lensed Lyman break galaxy MS~1512--cB58 at $z$ = 2.7276 (Pettini et al. 2002), thanks to the superior 
sensitivity of telescopes and spectrographs in the optical. Up to now only $FUSE$ spectra of the BCDs 
I~Zw~18 (Vidal-Madjar et al. 2000; Levshakov, Kegel, \& Agafonova 2001), Mrk~59 (Thuan, Lecavelier des 
Etangs, \& Izotov 2002), and NGC~1705 (Heckman et al. 2001) have already been published. However, I~Zw~18 
has been re-observed by $FUSE$ with an additional $\sim$~60,000 s of integration time (only $\sim$ 30,000 s 
are available for the first dataset) since the publication of the data. I~Zw~18 is also one of 
the most interesting and modelled BCDs in the local Universe. For all these reasons, we have retrieved 
from the archive, combined, and analyzed the two $FUSE$ datasets of I~Zw~18. This paper reports the 
results of our analysis. 

Discovered by Zwicky in 1966, I~Zw~18 remains the BCD with the lowest 
metallicity inferred from the oxygen content of its H~{\sc ii} regions (1/50 Z$_{\odot}$; e.g., Izotov 
\& Thuan 1999). Although it is often referred to as ``the most metal-poor galaxy known'', it is still 
two orders of magnitude more metal rich than the most metal-poor stars in the Milky Way (Fe/H $\simeq$ 
$-$\,4; Cayrel 1996). I~Zw~18 has always been regarded as the best candidate for a truly ``young'' 
galaxy  in the local Universe, with an estimated age less than 40 Myr (Izotov \& Thuan 1999). Its youth 
has, however, been questioned by the recent HST discovery in the optical (Aloisi, Tosi, \& Greggio 1999) 
and near-infrared (\"Ostlin 2000) of a well-defined population of asymptotic giant branch (AGB) stars at 
least several hundred Myr and possibly up to a few Gyr old. Moreover, detailed modeling of broad-band 
colors in the optical and near infrared (Hunt, Thuan, \& Izotov 2003) has shown that ages of a few 
hundred Myr are in better agreement with the integrated properties of I~Zw~18, even if as much as 22\% 
of the total mass could be contributed by older stars. Due to its very peculiar nature, I~Zw~18 has been 
extensively studied for the last three decades, but its evolutionary state still remains a matter of 
debate.

The distribution and kinematics of I~Zw~18 neutral hydrogen from aperture-synthesis observations have 
been discussed by several authors. The dynamical and H~{\sc i} masses are $\lesssim$ 10$^9$ M$_{\odot}$ 
and $\lesssim$ 10$^8$ M$_{\odot}$, respectively (Lequeux \& Viallefond 1980; Viallefond, Lequeux, \& 
Comte 1987; van Zee et al. 1998b). Van Zee et al. (1998b) have revealed a complex H~{\sc i} morphology 
with a neutral gas velocity dispersion of $\sigma \simeq$ 12$-$14 km s$^{-1}$. Martin (1996) and Petrosian 
et al. (1997) have discussed the complicated velocity field of the ionized component. The H~{\sc i} column 
density has been estimated to be as high as $N$(HI) $\simeq$ 2$-$3\,$\times\,10^{21}$ cm$^{-2}$ (Kunth et 
al. 1994; Vidal-Madjar et al. 2000). More recently, from their $STIS$ data at 0$\farcs$5 spatial resolution 
Brown et al. (2002) have discovered significant inhomogeneity in the neutral gas of I~Zw~18 with a peak 
$\sim$ 10 times higher than previously estimated. Molecular gas has not been detected in I~Zw~18, either 
through CO emission in the mm (Gondhalekar et al. 1998) or through H$_2$ absorption in the far-UV 
(Vidal-Madjar et al. 2000). Low extinction is reported in most studies, and the galaxy has not been detected 
by the Infra-Red Astronomical Satellite ($IRAS$). This could be an indication of a low and/or clumpy dust 
content in I~Zw~18 (Cannon et al. 2002), consistent with the low metallicity and $FUSE$ non-detection 
of diffuse H$_2$ (Vidal-Madjar et al. 2000).

Abundances derived from H~{\sc ii} regions are reasonably well known in I~Zw~18 (Searle \& Sargent 1972; 
Dufour, Garnett, \& Shield 1988; Skillman \& Kennicutt 1993; Garnett et al. 1995a, 1995b; Garnett et al. 
1997; Izotov \& Thuan 1999; Izotov et al. 1999). These studies indicate a low O content, but rather high 
N/O and C/O ratios (suggestive of an intermediate-age stellar population; e.g., Dufour et al. 1988), as
well as a high degree of uniformity in the nebular abundances. The metal content in the H~{\sc i} component 
of I~Zw~18 is instead more uncertain. Kunth et al. (1994) have attempted a first measurement of O in the 
neutral gas phase, and found a metallicity 20 times lower than in the H~{\sc ii} regions. Pettini \& Lipman 
(1995) have, however, shown that the observations of Kunth et al. cannot constrain the metallicity of the 
neutral gas due to saturation of the UV absorption line used, O~{\sc i} $\lambda$1302. Combining the analysis 
of Pettini \& Lipman (1995) with their new value for the H~{\sc i} dispersion, van Zee et al. (1998b) have 
intsead derived an H~{\sc i} metallicity more similar to that in the H~{\sc ii} regions. Also Levshakov et 
al. (2001) have recently suggested a good degree of mixing between warm and cold gaseous phases. At the 
opposite extreme, Izotov, Schaerer, \& Charbonnel (2001) have claimed primordial abundances for the neutral
gas in I~Zw~18 by investigating the possibility that UV absorption lines arise from H~{\sc ii} regions.

Our new $FUSE$ observations will certainly help to address many of the abundance issues in I~Zw~18. These 
data represent an improvement over $FUSE$ spectra previously published by Vidal-Madjar et al. (2000) and 
used for metal abundance determinations by Levshakov et al. (2001). The signal-to-noise ratio (S/N) is 
higher by a factor of 1.7, and the covered spectral range is wider ($\lambda\lambda$ = 917$-$1188 \AA~ 
{\it versus} $\lambda\lambda$ = 979$-$1188 \AA). We are able to estimate abundances of O, Si, Ar, Fe and 
N in the neutral gas of I~Zw~18 with unprecedented accuracy, thanks to the availability of several lines 
per ion and the application of a procedure allowing for their simultaneous fit. We find that the metal 
content in the neutral gas is significantly lower than in the H~{\sc ii} regions, apart from Fe, which 
is the same. A certain amount of ancient star-formation is required to reproduce the observed abundance 
patterns. The paper is outlined as follows. We describe the $FUSE$ observations and data reduction in 
section~2. The analysis of $FUSE$ data and its caveats are presented in section~3. The problem of the 
systemic velocity is addressed in section~4. Section~5 presents an estimate of the H~{\sc i} column 
density. We derive a $3 \sigma$ upper limit on the amount of diffuse H$_2$ in section~6. Heavy element 
abundances of the neutral ISM are derived in section~7. Section~8 presents our findings in the context 
of the chemical evolutionary state of I~Zw~18. The summary and conclusions are given in section~9.

\section{$FUSE$ Observations and Data Reduction}

Two $FUSE$ datasets for I~Zw~18 ($\alpha_{2000} = 09^h\,34^m\,01\fs92$, 
$\delta_{2000} = +\,55\arcdeg\,14\arcmin\,26\farcs1$; $l_{II} = 160\fdg52610$, $b_{II}$ = +\,44\fdg84183) 
were obtained on 1999 November 27 and 2001 February 11, respectively, as part of$FUSE$ Team programs P198 
and P108. The first observations detected the target only in the LiF channels ($\lambda\lambda$ = 979$-$1188 
\AA) for a total exposure time of $\sim$ 31,600~s, $\sim$ 30\% of which occurred during the orbital night. 
These spectra have been extensively analyzed by Vidal-Madjar et al. (2000) and Levshakov et al. (2001). 
The most recent dataset includes both LiF and SiC channels ($\lambda\lambda$ = 905$-$1188 \AA) for a total 
integration time of $\sim$ 63,500~s, of which $\sim$ 60\% occurred during orbital night. In both observing
runs I~Zw~18 was centered in the large aperture (LWRS, $30\arcsec\times\,30\arcsec$) which amply covered the 
whole galaxy. 

The data were retrieved from the MAST archive and processed with the latest release of the $FUSE$ calibration 
pipeline (version 2.0). This was, however, updated to include the new set of background files and the correct 
error estimate that were later available in version 2.1\footnote{ It has been later discovered (FUSE User's 
Electronic Newsletter no.~25 of 2003 March 17) that the calculation of the errors in the FUSE calibration 
pipeline version 2.1 has a bug that could underestimate the real value of the errors for bright sources. This 
clearly does not affect faint targets like I~Zw~18.}. The first part of the pipeline was applied to individual 
exposures in order to get correct Doppler shifts and grating motions from the orbital parameters in the headers 
of single images. All exposures relative to each dataset were then combined with the CALFUSE task ``ttag\_combine'',
taking into account residual velocity offsets. We finally ran the last part of the calibration pipeline on the 
final combined raw images. This non-standard procedure allowed for a better estimate of the background and a more
accurate spectral extraction in case of faint sources. A binning by a factor of 6 was applied to the data in 
order to improve the identification of spectral features without degrading the spectral resolution. The extracted 
1D spectra, one per dataset, were then combined together to obtain a unique spectrum with an exposure time around 
95,100 s and 63,500 s in the LiF and SiC channels, respectively. This increased the S/N per resolution element 
by a factor of $\sim$ 1.7 in the LiF channels compared to the original spectra, and extended the spectral 
coverage to $\lambda$ $<$ 979~\AA~ in the SiC channels. Residual wavelength shifts among different segments 
were then estimated and applied using the LiF~1A channel as a reference. A remaining zero-point wavelength 
offset of $\lesssim 0.10$ \AA~ was corrected by requiring the Galactic ISM lines to fall at $\sim$ 0 km s$^{-1}$. 
  
We avoided combining data from the eight different channels into one composite spectrum over the whole 
$FUSE$ spectral range because the instrumental resolution and sensitivity vary as a function of wavelength. 
We instead constructed a final spectrum in the interval $\lambda\lambda$ = 917$-$1188 \AA~ by adding together 
segments of spectra taken from the channel with the highest sensitivity in a certain wavelength region, mainly 
SiC~2A, LiF~1A, and LiF~2A. The redundant information at lower S/N contained in the segments not used was, 
however, considered for a consistency check.

Only data taken during the orbital night (with a S/N lower by a factor of 1.3) were considered in those 
parts heavily contaminated by emission from terrestrial O~{\sc i} and N~{\sc i} airglow. This procedure 
was applied in order to minimize geocoronal contamination. We expected emission from geocoronal lines 
to be practically absent in the so-called ``night'' data. This can be easily verified by an eye inspection
of the final spectrum, where, e.g., the saturated O~{\sc i} $\lambda$1039 absorption from the Milky Way 
goes to zero intensity, indicating practically no residual contribution from one of the strongest airglow 
lines. On the other hand, absorption lines in I~Zw~18 are redshifted of a few pixels (e.g., $\sim$ 2.5 
\AA~for O~{\sc i} $\lambda$1039), so that residual airglow contamination does not represent a problem 
in any case.

A nominal spectral resolution of $\Delta\lambda$/$\lambda$ $\simeq$ 10000, corresponding 
to $v$ $\simeq$ 30 km s$^{-1}$, was derived by taking into account the aperture and the size 
of the galaxy (10\arcsec). An observed upper limit around 40 km s$^{-1}$ was instead estimated 
by measuring the FWHM of the narrowest H$_2$ lines arising from the Milky Way (MW). A final 
spectral resolution of 35 $\pm$ 5 km s$^{-1}$ over the whole wavelength range was finally 
adopted. The noise spectrum was calculated through photon statistics propagation of object 
and sky spectra. The final S/N is $\sim$ 7, 17, and 18 per resolution element at 950 \AA~ 
(SiC~2A), 1050 \AA~ (LiF~1A), and 1150 \AA~ (LiF~2A). 

The fit of the continuum was performed by interpolating between points of the observed flux 
free of apparent absorption. This is not particularly easy because of the limited S/N of the 
spectrum. Members of the team independently estimated the continuum for a consistency check, 
and the differences were always well below the noise uncertainties. Errors associated with 
measured quantities in the $FUSE$ spectrum are from counting statistics only and were calculated 
from the S/N of the data. Systematic errors due to the uncertainty in the continuum placement 
were not considered.

The resulting final $FUSE$ spectrum of I~Zw~18 is shown in Fig.~\ref{compspec}. Three absorption 
line systems at three different radial velocities are clearly present: the well-known high-velocity 
cloud (HVC) at $-$\,160 km s$^{-1}$, the Milky Way (MW) at 0 km s$^{-1}$, and I~Zw~18 at $\sim$ 750 
km s$^{-1}$. The high-velocity cloud and Galaxy systems have already been briefly discussed by 
Vidal-Madjar et al. (2000), and will not be detailed further. Here we will concentrate on the 
absorption line system in I~Zw~18.

\section{Analysis of $FUSE$ Data and Caveats}

We derived column densities of H~{\sc i} and heavy elements from the $FUSE$ spectra of 
I~Zw~18 by line-profile fitting of the observed absorption lines. Standard theoretical 
Voigt profiles were convolved with the gaussian instrumental resolution and fitted to 
the data. The package FITLYMAN (Fontana \& Ballester 1995) in MIDAS for multi-component 
fitting was used for this purpose. This method is more powerful than a simple curve-of-growth 
analysis (based on the equivalent width of the absorption lines) because it allows for $i)$ 
the deblending of multiple components along the line of sight contributing to the same 
absorption, and $ii)$ the simultaneous and independent fit of contaminating absorption 
components. We were able to fit absorption lines of all the ions considered in I~Zw~18 
with very simple velocity component models (one component, except for Fe~{\sc ii} where 
two components were necessary).

The real physical situation described by I~Zw~18 far-UV spectra is, however, more complex. $FUSE$ 
detects a non-linear average absorption over the full extent of the stellar background sources, 
since its large aperture includes the whole galaxy. This implies that the observed lines arise 
from a combination of many unresolved velocity components from different absorbing clouds along 
the many lines of sight. Moreover, some lines of sight may have saturated absorption, even if the 
composite profile does not go to zero intensity (e.g., Savage \& Sembach 1991).

Jenkins (1986), on the other hand, has demonstrated that the single-velocity approximation 
(one-velocity component) applied to complex blends of features gives nearly the correct 
answer (the simulated-to-true column density ratio rarely goes below 0.8) if the distribution 
function for the line characteristics is not irregular (e.g., bimodal, see also Savage \& 
Sembach 1991). This result holds also if different lines have different saturation levels or 
Doppler parameters $b$. Since our $FUSE$ data of I~Zw~18 average over the whole galaxy, we 
expect a quite regular distribution of the kinematical properties of the single absorbing 
components. We thus believe we amply fall within the regime were the single-velocity 
approximation is valid.

In light of these considerations, we preferred to maintain a simple approach in the 
determination of the column densities with the line-profile fitting method. We thus 
avoided the introduction of additional free parameters, i.e. the number of intervening 
clouds and their velocity distribution, since we believe the resolution of the data does
not allow to correctly constrain the real physical situation represented by this type of 
observation. In the single-velocity approximation the fitting parameter $b$ has no precise 
physical meaning, but is rather the result of the combination of both the various line 
Doppler widths present (due, e.g., to turbulent and/or thermal broadening) and the various 
velocity separations among the different line components (Hobbs 1974). On the other hand, 
according to Jenkins (1986) the column density is well constrained. In addition, the column
density of a certain ion is even better constrained if several lines with different values 
of $f \lambda$ ($\lambda$ is the rest-frame wavelength and $f$ the oscillator strength of 
the absorption) are available for a simultaneous fit, and the results are independent of 
saturation problems affecting the strongest lines.

However, in order to check the quality of the results with the line-profile fitting and 
the single-velocity approximation, we also determined total column densities of heavy 
elements by applying the apparent optical depth method (Savage \& Sembach 1991). The 
apparent column density of an ion in each velocity bin, $N_{\rm a}(v)$ in units of cm$^{-2}$ 
(km s$^{-1}$)$^{-1}$, is related to the apparent optical depth in that bin $\tau_{\rm a}(v)$ 
by the expression:
\begin{equation}
N_{\rm a}(v) ~=~  \frac{m_e\,c}{\pi\,e^2} ~ \frac{\tau_{\rm a}(v)}{f\lambda}
~=~  3.768\,\times\,10^{14} \frac{\tau_{\rm a}(v)}{f\lambda{\rm (\AA)}}~.  
\end{equation}
The apparent optical depth is directly calculated from the observed intensity in the line at 
velocity $v$, $I_{\rm obs}(v)$, by 
\begin{equation}
\tau_{\rm a}(v) ~=~ - \ln~ [I_{\rm obs}(v)/I_0(v)]~,
\end{equation}
where $I_0(v)$ is the intensity in the continuum. $N_{\rm a}(v)$ is an apparent column density 
per unit velocity because its value depends on the resolution of the spectrograph and on the 
apparent shape of the line. The total apparent column density, $N_{\rm a}$, is obtained by 
direct integration of equation (1) over the velocity interval where line absorption takes 
place
\begin{equation}
N_{\rm a} ~=~ \int N_{\rm a}(v)\,d\,v~.
\end{equation}
In the limit where the absorption line is weak ($\tau$ $\ll$ 1) or fully resolved (the FWHM of 
the line is larger than the instrumental FWHM), the total apparent column density $N_{\rm a}$ 
and the true column density $N$ are equal. 

The apparent optical depth method is very powerful for determining column densities. Its 
strength lies in the fact that no assumption needs to be made concerning the velocity 
distribution of absorbers, i.e., it does not depend on the number of intervening clouds 
along the single or multiple lines of sight. However, it does not provide a correct answer 
for quite strong lines that are not fully resolved. In the case of our resolution of $\sim$ 
35 km s$^{-1}$, the line-profile fitting should give a more accurate measure of the ionic 
column densities compared to the apparent optical depth method, especially for stronger 
lines. The latter can nevertheless be more potent in unveiling hidden saturation if two 
or more lines with relatively different $f\lambda$ values are available. Such a situation 
will easily manifest itself with a column density from the apparent optical depth method 
being smaller for lines with higher $f\lambda$ values. In our analysis, the similarities 
between the column densities obtained with the line-profile fitting and the apparent optical 
depth method for weak (and thus, more optically thin) lines ($W_{\rm 0} \lesssim 200$ m\AA) 
strengthen our results, and justify the single-velocity approximation approach (\S~7.1). 

Finally, we want to point out another caveat generic to this type of $FUSE$ observations where
the whole star-forming galaxy is included in the spectroscopic aperture. The far-UV radiation 
we detect is the one not intercepting the opaque dense clumps with a high dust and molecular 
content (even if with a very low covering factor; Hoopes et al. 2003b). The far-UV light is 
thus biased toward lower metallicity regions. The more diffuse ISM, however, can contain a
certain amount of dust (Meurer, Heckman, \& Calzetti 1999; Hoopes et al. 2003b), thus of metals. 
In addition, there are indications from UV/optical studies that the dust content in I~Zw~18 is 
relatively low (Meurer et al. 1999; Cannon et al. 2002). We hence do not expect that our results 
are heavily biased toward regions with lower metallicity. Nevertheless, the strongest result 
of our analysis, i.e. the metal enrichment of the neutral gas, would still remain valid even in 
case dust plays an important role.

\section{The Systemic Velocity of I~Zw~18}

The $FUSE$ spectrum of I~Zw~18 exhibits a multitude of stellar and interstellar lines. Most of 
the stellar lines are blends of multiple transitions. The strongest and cleanest photospheric 
line is C~{\sc iii} $\lambda$1176, although this is a multiplet centered at 1175.6 \AA. We observed 
this line at a central wavelength of $\lambda$ = 1178.59 $\pm$ 0.04 \AA, taking into account 
that its red side could be contaminated by a P-Cygni profile from stellar winds of OB supergiants. 
We inferred a redshift of $z_{\rm stars}$ = 0.00254 $\pm$ 0.00003, corresponding to a systemic 
velocity of $v_{\rm stars}$ = 761 $\pm$ 9 km s$^{-1}$. Exactly the same result is independently 
obtained for the other strong C~{\sc iii} line at 977 \AA, whose nature is more uncertain since 
it can arise from both the photosphere and the interstellar medium. Our value of $v_{\rm stars}$
is very similar to the mean value of $v_{\rm HII}$ $\simeq$ 763 km s$^{-1}$, obtained by Petrosian et 
al. (1997) from an H$\alpha$ interferometric study of both the northwest and southeast H~{\sc ii} 
regions ($v_{\rm HII}$ = 742 $\pm$ 7 km s$^{-1}$ and 783 $\pm$ 5 km s$^{-1}$, respectively). Within 
the errors, $v_{\rm stars}$ is also consistent with $v_{\rm HI}$ = 751 $\pm$ 2 km s$^{-1}$ derived 
by Thuan et al. (1999) for the H~{\sc i} component. Here we will adopt $z_{\rm stars}$ = 0.00254 as 
the redshift of I~Zw~18 and $v_{\rm stars}$ = 761 km s$^{-1}$ as its systemic velocity.

\section{H~{\sc i} Column Density}

The good S/N and high resolution of the $FUSE$ spectrum of  I~Zw~18 give us, in principle, a unique 
opportunity to study the absorption lines of the H~{\sc i} Lyman series from Ly\,$\beta$ to 
Ly\,$\mu$ (H~{\sc i}~12). In practice only the absorptions up to Ly\,$\eta$ (H~{\sc i}~7) were 
used, because the higher order lines were completely contaminated by Galactic H$_2$ and other 
interstellar lines. Ly\,$\alpha$ at $\lambda$ = 1215.67 \AA~ is outside the covered spectral range 
(Fig.~\ref{compspec}). The foreground H~{\sc i} column density in the direction of I~Zw~18 due to 
the Milky Way is $\sim$ $2\,\times\,10^{20}$ cm$^{-2}$ (Stark et al. 1992), while the high-velocity 
cloud contribution is $\sim$ $2.1\,\times\,10^{19}$ cm$^{-2}$ (Kunth et al. 1994). 

The H~{\sc i} absorption lines of I~Zw~18 appear to be narrow, and thus interstellar in origin. However, 
$FUSE$ spectra of Galactic early B stars (Pellerin et al. 2002) clearly show the presence of a broad 
photospheric component (e.g., their Fig.~3). This implies that when early B stars start to dominate the
integrated spectrum of a stellar population (e.g., a burst with an age greater than $\sim$ 10 Myr), the 
wings of a large photospheric contribution are superposed on the cores of the narrow interstellar 
absorption. The same conclusion is reached by Gonz\'alez-Delgado, Leitherer, \& Heckman (1997) through 
their comparison of observed and synthetic profiles of O~VI $\lambda\lambda$1032, 1038 + Ly\,$\beta$ + 
C~II $\lambda\lambda$1036, 1037 (see their Fig.~4). 

The age of the stellar population dominating the emission-line UV/optical spectra of I~Zw~18 is quite 
well defined. De Mello et al. (1998) have derived spectral information on the northwest star-forming 
region from the literature, and dated its stellar content with stellar evolutionary synthesis models 
at a metallicity of Z\,=\,0.0004. They have suggested an instantaneous burst with an age around 3 Myr, 
as indicated by the observed Wolf-Rayet stellar features and the equivalent width of H$\beta$. On the 
other hand, Mas-Hesse \& Kunth (1999) have inferred the age of I~Zw~18 from ground-based optical 
spectroscopy of the whole galaxy. The evolutionary synthesis models they have applied present a 
degeneracy for the best solution: a continuous star formation with an age of 13 Myr (this model is 
preferred by these authors), and a 3 Myr old instantaneous burst. Mas-Hesse \& Kunth have also been 
able to assess that the underlying older stellar component, (as indicated by studies on I~Zw~18 
resolved stellar population), would not affect the dating of the observed spectrum.

At present it is not possible to determine the age of I~Zw~18 stellar population by applying 
stellar evolutionary synthesis codes to the $FUSE$ data. We have recently implemented a $FUSE$ 
stellar library of stars from the Milky Way and Magellanic Clouds into Starburst99 (Robert et 
al. 2003). This implies that the synthesis of stellar populations in the spectral range $\sim$ 
1003$-$1188 \AA~ is available, but still limited to solar and 1/5 solar metallicities. We are 
investigating the possibility of implementing Starburst99 with a stellar library of theoretical 
stellar atmosphere spectra at extremely low metallicities (Kudritzki 2002). This will allow us 
to correctly date I~Zw~18 stars in the far-UV band in the near future. In this paper we will 
assume that the $FUSE$ spectrum is dominated by hot O stars from star-forming regions (as 
previously suggested by other authors). The additional presence of B or later-type stars 
would not affect an O-dominated far-UV spectrum due to the much lower luminosities contributed 
by these stars; e.g., a late-type O star with $T \simeq 30,000$ K at 1100 \AA~has a luminosity 
which is $\sim$ 100 times higher than that of a middle-type B star with $T \simeq 15,000$ K,
and this difference is even larger if later-type stars are considered (Kurucz 1979). We thus 
expect that the wings of H~{\sc i} lines are not contaminated by broad photospheric absorption 
in I~Zw1~18. A single narrow interstellar component has been also considered by Vidal-Madjar et 
al. (2000) for the determination of H~{\sc i} column density from their $FUSE$ spectra of I~Zw~18. 
Moreover, in their analysis of $FUSE$ data of Mrk~59, Thuan et al. (2002) explicitly mention 
contamination of H~{\sc i} Lyman series lines by broad photospheric absorption. They attribute 
the difference in photospheric contamination between I~Zw~18 and Mrk~59 to the older age of the latter.

We reproduced the shape of the H~{\sc i} Lyman series lines in I~Zw~18 through profiles characterized 
by a single component with a column density of $N({\rm HI})$ = $2.2^{+\,0.6}_{-\,0.5}\,\times\,10^{21}$ 
cm$^{-2}$. Fig.~\ref{lymanplot} shows the Lyman series lines with the best theoretical model. Only the 
red wing of Ly$\beta$ was used to constrain the model parameters, since the blue wing is contaminated 
by Ly$\beta$ absorption from the Galaxy and the high-velocity cloud, as well as by geocoronal emission.
Due to its damped profile, Ly$\beta$ is a very strong constraint to the H~{\sc i} column density. This 
is valid also in our case, where the Lyman series lines of higher order show some contamination by 
interstellar absorptions, except for L$\eta$. On the other hand, L$\eta$ is saturated but not damped, 
thus its profile is much more sensitive to the Doppler parameter.

Our H~{\sc i} column density is perfectly consistent with the value of $2.1\,\times\,10^{21}$ cm$^{-2}$ 
that Vidal-Madjar et al. (2000) obtained from Ly$\beta$ absorption in their $FUSE$ spectrum. It is also 
similar to, but slightly less than, the value of $3.5\,(\pm\,0.5)\,\times\,10^{21}$ cm$^{-2}$ inferred 
by Kunth et al. (1994) from Ly$\alpha$ absorption. In addition, the systemic velocity of $v_{\rm HI}$ 
= 753 $\pm$ 6 km s$^{-1}$ ($z_{\rm HI}$ = 0.00251 $\pm$ 0.00002) we inferred for the H~{\sc i} gaseous 
component, is compatible with the value derived by Thuan et al. (1999) from 21-cm observations 
($v_{\rm HI}$ = 751 $\pm$ 2 km s$^{-1}$). 

Brown et al. (2002) have recently obtained $STIS$ data of I~Zw~18 at 0$\farcs$5 spatial resolution 
and inferred the H~{\sc i} column density from Ly$\alpha$ absorption at different locations within 
the galaxy. They have discovered significant inhomogeneity in the neutral gas, with a peak as high 
as $N({\rm HI})$ $\simeq$ $2\,\times\,10^{22}$ cm$^{-2}$ in the fainter (southeast) part of the galaxy, 
and a density quickly dropping to a value 10 times lower as soon as a slightly different line of sight 
is considered. Our $FUSE$ estimate of $N({\rm HI})$ is $\sim$ 10 times lower than the peak value measured 
by Brown et al., but the difference can be easily ascribed to the different nature of the observations. 
$FUSE$ data have been acquired through an aperture much larger than the whole extension of I~Zw~18. 
$STIS$ data are instead a collection of 7 different long-slit observations covering a large fraction of 
the galaxy, but not its totality. As a consequence, our data are clearly a non-linear average value of 
the H~{\sc i} column density over many lines of sight through the galaxy, and are probably consistent 
with the Brown et al. results, once smearing of the higher spatial resolution information is correctly 
taken into account.

The Doppler parameter for the best theoretical model of the H~{\sc i} absorption lines is $b$ = 35 
$\pm$ 10 km s$^{-1}$. Taking into account that $FUSE$ data are sampling an average absorption (\S~3), 
this value of $b$ does not have a real physical meaning, being associated with unresolved multiple 
velocity components along the various lines of sight. For comparison, the instrumental broadening
due to our spectral resolution is around 21 km s$^{-1}$, too low to resolve the various velocity 
components with typical temperatures $T \lesssim 10^4$ K if they are dominated by thermal broadening
(in this case $b \lesssim 10$ km s$^{-1}$). On the other hand, the velocity dispersion of $\sigma$
$\simeq$ 12$-$14 km s$^{-1}$, as inferred from radio observations (van Zee et al. 1998b), would lead
to a value of $\sim$ 20 km s$^{-1}$ ($b$ = $\sqrt{2}\sigma$) for single velocity components where 
broadening is dominated by turbulent motions. It is interesting to notice that there is a velocity 
gradient of about 50 km s$^{-1}$ in the H~{\sc i} component across the area covered by the $FUSE$ 
beam (van Zee et al. 1998b). It could be that most of our large Doppler parameter is due to this 
gradient, since the far-UV light from the background stars probably samples a large fraction of this 
range in H~{\sc i} velocity.

\section{Upper Limits on the Diffuse H$_2$ Content}

H$_2$ absorption lines from the Milky Way are clearly visible in the  I~Zw~18 spectrum (Fig.~1). 
However, no lines of H$_2$ are seen at the radial velocity of the BCD, despite the fact that 
I~Zw~18 has a total H~{\sc i} column density higher than that measured on average in the Milky Way. 
Vidal-Madjar et al. (2000) estimated a 10\,$\sigma$ upper limit of 10$^{15}$ cm$^{-2}$ for the 
molecular hydrogen column density. In this paper we re-addressed this issue by taking advantage 
of the higher S/N $FUSE$ spectrum.

We inferred an upper limit to the H$_2$ column density of I~Zw~18 with the following procedure.
For each of the first five ($J$ = 0$-$4) rotational levels of H$_2$ intrinsic to I~Zw~18 we
selected the strongest unblended line providing the most stringent constraint to the inferred 
column density in each level, we measured the noise at its predicted location and we calculated 
an upper limit to its rest-frame equivalent width $W_0$. The transitions considered for this purpose 
are indicated in Table~1 together with their $W_0$ values. Since we were dealing with upper limits 
for the equivalent width, we could not apply the curve-of-growth technique and estimate a total 
H$_2$ column density. Therefore, we converted the $W_0$ upper limits of each transition into column 
density upper limits by assuming the optically thin case, corresponding to the linear part of the 
curve of growth (Spitzer 1978). The derived values of $N_{\rm J}$ are listed in Table~1. 

The absorption lines of the $J \ge 2$ levels are intrinsically weaker than those of the $J = 0$ and
$J = 1$ levels, resulting in less restrictive upper limits on the column densities of those levels. 
Since the bulk of the column density is contained in the lower levels, deriving a limit on total 
H$_2$ by simply summing the upper limits for all the levels will be weighted toward the higher $J$ 
levels and thus will produce an unrealistically high upper limit. We can improve the constraints on
total H$_2$ if we take into account the likely distribution of the level populations.

Level populations for H$_2$ are typically described by a two-temperature Boltzmann distribution (see, 
e.g., Sembach et al. 2001). The kinetic temperature is usually suitable to describe the situation for 
$J$ = 0 and $J$ = 1 levels, since the densities are high enough that collisions determine the level 
populations. The populations of higher levels can be affected by other processes, like UV photon pumping, 
shocks, and formation of H$_2$ on dust grains (see Shull \& Beckwith 1982 and references therein), and 
are usually better described by a higher excitation temperature. In both cases, the temperatures are 
derived from the ratios of column densities $N_{\rm J}$ and statistical weights $g_{\rm J}$ following 
Spitzer, Cochran, \& Hirshfeld (1974).

Our procedure for setting an upper limit on total H$_2$ column density is as follows. For the $J = 0$ 
and $J = 1$ levels we simply add the two measured upper limits. For the higher $J$ levels we assume a 
temperature, which establishes the level populations, and then normalize the level populations so that 
they do not violate any of the measured upper limits. Figure~\ref{H2intr} shows how this is done in an 
excitation diagram. The $3 \sigma$ upper limit to the column density $N_{\rm J}$ of each $J$ level, 
divided by its statistical weight $g_{\rm J}$, is plotted against excitation energy $E_{\rm J}$. One 
of the two dashed lines in Fig.~\ref{H2intr} shows the prediction for the extreme case of an excitation 
temperature of $T$ = 1000 K. While this temperature is quite high for interstellar H$_2$, comparable 
temperatures have been observed in supernova remnants (e.g., Welsh, Rachford, \& Tumlinson 2002), so 
it may be applicable to starburst regions. Lower temperatures would give lower total column densities,
so adopting 1000 K gives a conservative upper limit. The level populations are normalized so that they 
do not violate any of the observed upper limits, which means the total column density for the higher 
levels is dictated by the $J$ = 1 upper limit. Note that this assumed temperature is used only to 
estimate the level populations, and that we cannot measure the excitation temperature in the H$_2$ 
gas without detections of H$_2$ in the individual $J$ levels.

We find a conservative $3 \sigma$ upper limit of log\,$N$(H$_2$) $\lesssim$ 14.72 cm$^{-2}$ by adding 
the observed upper limits for the $J$ = 0 and $J$ = 1 levels to the theoretical upper limits (which 
are lower than the measured ones) calculated for the $J \ge$ 2 levels from the $T$ = 1000 K distribution. 
If the higher $J$ levels were described by a more standard $T$ = 500 K distribution (the other dashed 
line in Fig.~\ref{H2intr}), the upper limit on the column density would be lower and equal to log\,$N$(H$_2$) 
$\lesssim$ 14.55 cm$^{-2}$.

We derive a molecular hydrogen fraction of $f_{\rm H_2}$ = 2$N$(H$_2$)/[$N$(HI) + 2$N$(H$_2$)] $<$ 
4.7 $\times$ 10$^{-7}$ by assuming the previously estimated values of $N$(HI) = 2.2\,$\times$\,10$^{21}$ 
cm$^{-2}$ and $N$(H$_2$) $<$ 5.25 $\times$ 10$^{14}$ cm$^{-2}$. The mass in H~{\sc i} of the cloud 
associated with the optical body of I~Zw~18 (and probably completely included in the $FUSE$ aperture) 
is $\sim$ 2.6 $\times$ 10$^7$ M$_{\odot}$ (van Zee et al. 1998b). This translates into an upper limit 
of $\sim$ 12 M$_{\odot}$ for the total mass of diffuse H$_2$ in front of the sources of far-UV light 
in the galaxy. We cannot rule out that large amounts of clumpy molecular gas are present in I~Zw~18. 
Clumpy H$_2$ has probably the same spatial distribution of dust (e.g., Cannon et al. 2002), considering 
that the major mechanism of H$_2$ formation is on the surface of dust grains. We do not simply see 
clumpy H$_2$ because the $FUSE$ band selectively detects far-UV radiation passing through regions 
devoid of H$_2$ or intense enough to destroy H$_2$ molecules along the line of sight (Hoopes et al. 
2003b).

\section{Heavy Element Abundances}

\subsection{Column Densities}

\begin{figure}
\plotone{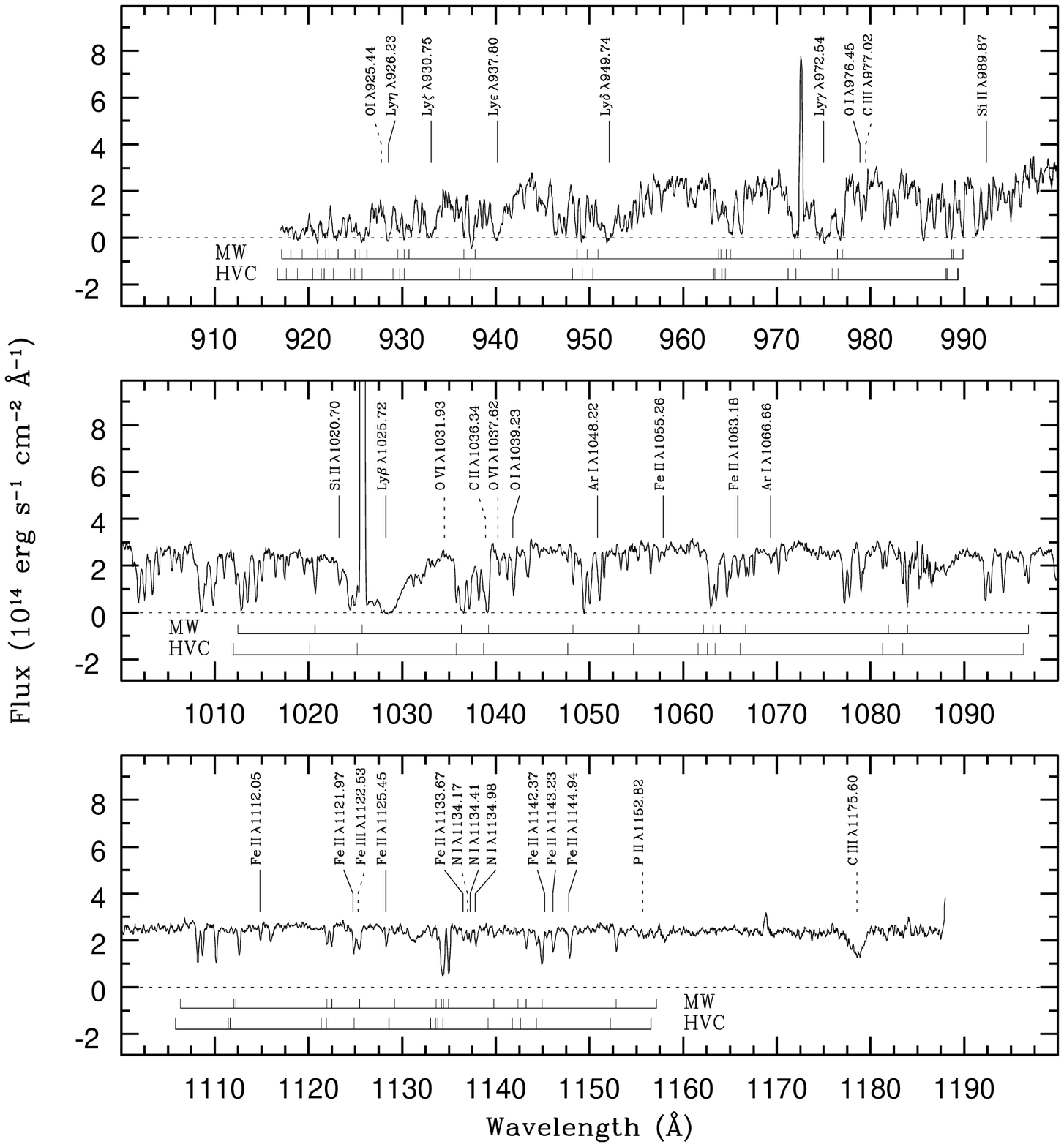}
\caption{I~Zw~18 $FUSE$ spectrum in the range 917$-$1188 \AA. The data have been 
smoothed by a 5-point boxcar. Uncontaminated interstellar absorption lines at 
$v$ $\simeq$ 750 km s$^{-1}$ used in the determination of metal abundances are 
indicated with solid lines. Other interstellar/stellar absorptions associated 
with the galaxy and discussed in our analysis are marked with dotted lines. The 
ticks indicate the position of all low-ionization interstellar absorption lines 
(see Prochaska et al. 2001 for a compilation) that could potentially arise from 
the Milky Way (MW) at $v$ = 0 km s$^{-1}$ or the well known high-velocity cloud 
(HVC) at $v$ = $-$\,160 km s$^{-1}$. Unmarked strong absorption lines below 1120 
\AA~ are due to molecular hydrogen from the ISM of the Milky Way (I~Zw~18 does 
not show detectable H$_2$).
\label{compspec}}
\end{figure}

\begin{figure}
\plotone{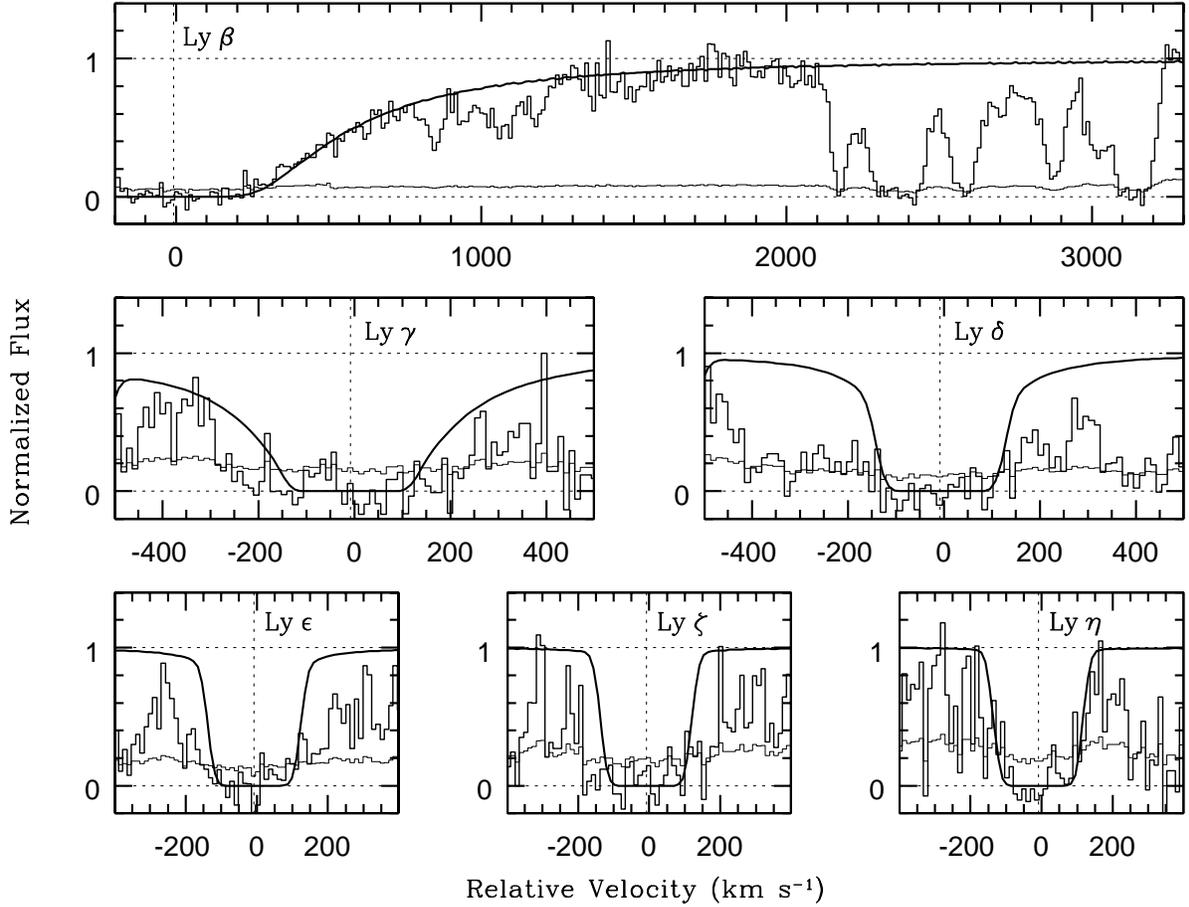}
\caption{H~{\sc i} Lyman series profiles in I~Zw~18. Observed spectrum ({\it histogram})
and theoretical line profile for $N$ = 2.2 $\times$ 10$^{21}$ cm$^{-2}$ and $b$ = 35 km s$^{-1}$ 
at $v$ = 753 km s$^{-1}$ ({\it solid line}). The velocity scale is relative to $v_{\rm stars}$ 
= 761 km s$^{-1}$. The spectrum of the noise per pixel is also shown ({\it thin line histogram}).
\label{lymanplot}}
\end{figure}

\begin{figure}
\plotone{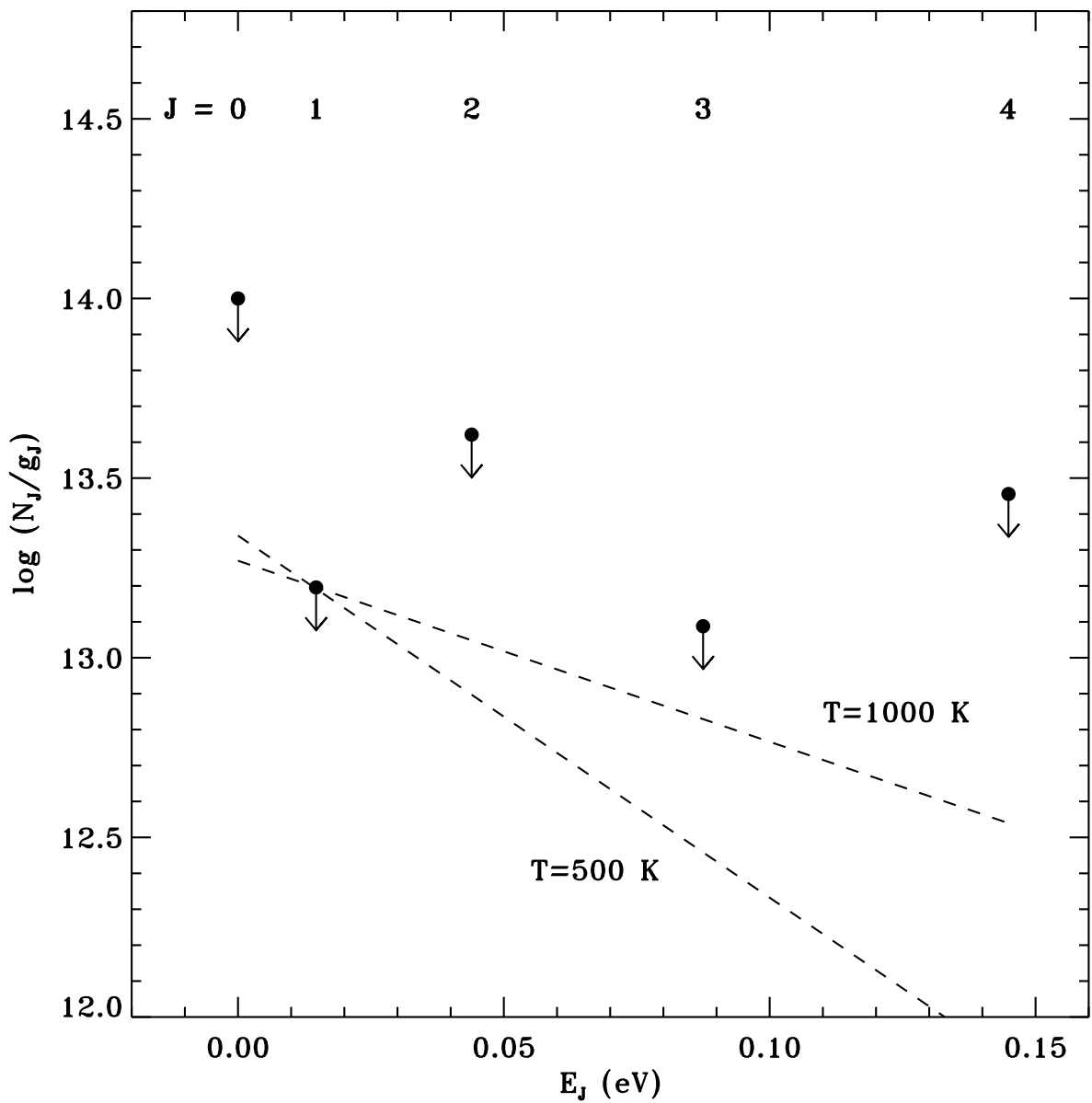}
\caption{Excitation diagram for H$_2$ intrinsic to I~Zw~18. The column density $N_{\rm J}$
divided by the statistical weight $g_{\rm J}$ is plotted {\it versus} the excitation potential 
$E_{\rm J}$ for each $J$ level. The points are $3 \sigma$ upper limits on the H$_2$ column 
densities derived from the strongest unblended transition for each of the first 5 rotational 
levels. The dashed lines are Boltzmann distributions for excitation temperatures of $T$ = 500 
and 1000 K. These are {\it assumed} temperatures used to improve the upper limit on the total 
H$_2$ column density. We cannot constrain the excitation temperature with the upper limits we 
have measured.
\label{H2intr}}
\end{figure}

Table~2 lists interstellar absorption lines measured in the $FUSE$ spectrum of I~Zw~18 (see also 
Fig.~\ref{compspec}). We cover several transitions of neutral and singly ionized atoms of heavy 
elements. Vacuum rest wavelengths $\lambda_{lab}$ (column 2) of the transitions are from the 
compilation by Morton (1991). Oscillator strengths $f$ (column 3) are from the references indicated 
in column 4. Rest-frame equivalent widths, $W_0$, and their 1\,$\sigma$ errors are listed in column 
10. In a few cases where an interstellar line was found to be blended with another unrelated feature, 
its equivalent width is a lower limit that does not include the blend.

We derived values of column density for ions of interest by using the line-profile fitting technique 
(\S~3). The line profiles of most ions appear symmetric, and were simultaneously fitted by a single 
velocity component. The only exception is Fe~{\sc ii}, where the asymmetric structure of its profile 
required the introduction of a second component in order to get an acceptable value for $\chi^2$. It 
is possible that an additional velocity component is well constrained for the Fe~{\sc ii} lines only
for the following reasons: {\it i)} there are many lines to consider for the fit; {\it ii)} the lines 
are quite faint, and the blend of the two components is less severe; {\it iii)} they are all located 
in the reddest part of the spectrum where the S/N is better. On the other hand, the fit with one 
component would give a similar iron column density within the uncertainties. In Table~2 we give 
the best-fit results for the ions considered: the redshift $z$ is listed in column 5, the Doppler width 
$b$ is reported in column 6, and the logarithm of the column density $N_{PF}$ is indicated in column 
7. In the case of Fe~{\sc ii}, we separately list the best fit parameters of both velocity components, 
as well as the total column density. 

\begin{figure}
\plotone{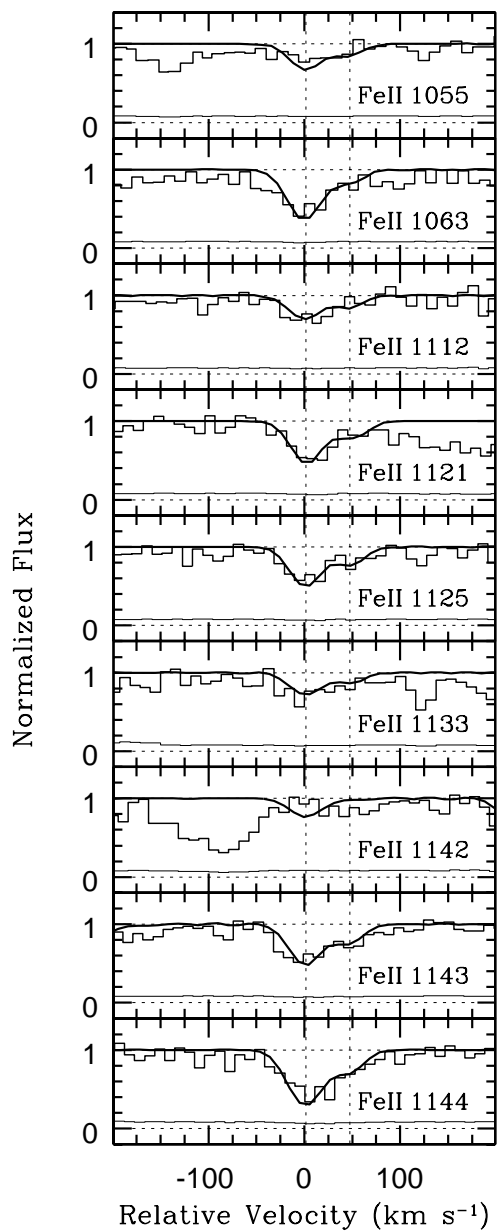}
\caption{Fe~{\sc ii} profiles in I~Zw~18: observed spectrum ({\it histogram}) and
theoretical line profile ({\it solid line}). A two-velocity component model has
been used: $N_1$ = 1.0 $\times$ 10$^{15}$ cm$^{-2}$ and $b_1$ = 8.8 km s$^{-1}$
at $v_1$ = 764 km s$^{-1}$ for the first component; $N_2$ = 2.3 $\times$ 10$^{14}$ 
cm$^{-2}$ and $b_2$ = 3.2 km s$^{-1}$ at $v_2$ = 809 km s$^{-1}$ for the second component. 
The velocity scale is relative to $v_{\rm stars}$ = 761 km s$^{-1}$. The spectrum of the 
noise per pixel is also shown ({\it thin line histogram}).
\label{FeIIplot}}
\end{figure}

\begin{figure}
\plotone{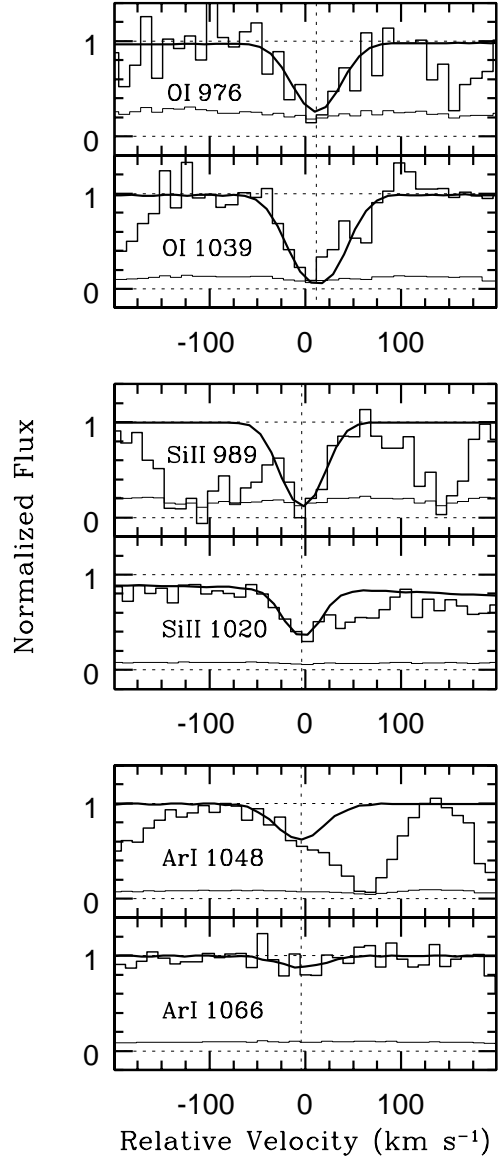}
\caption{Same as Fig.~\ref{FeIIplot}, but for the $\alpha$-elements. A single-velocity 
component has been fitted to the data with the following parameters: $N$ = 9.5 $\times$ 
10$^{15}$ cm$^{-2}$ and $b$ = 23 km s$^{-1}$ at $v$ = 774 km s$^{-1}$ for O~{\sc i} (top 
panel); $N$ = 6.5 $\times$ 10$^{14}$ cm$^{-2}$ and $b$ = 15.7 km s$^{-1}$ at $v$ = 758 km 
s$^{-1}$ for Si~{\sc ii} (middle panel); $N$ = 4.0 $\times$ 10$^{13}$ cm$^{-2}$ and $b$ = 
27 km s$^{-1}$ at $v$ = 757 km s$^{-1}$ for Ar~{\sc i} (bottom panel). The absorptions
contaminating Si~{\sc ii} $\lambda$1020 and Ar~{\sc i} $\lambda$1048 are from $H_2$ in the 
ISM of the Milky Way and have been independently modelled.
\label{alphaplot}}
\end{figure}

\begin{figure}
\plotone{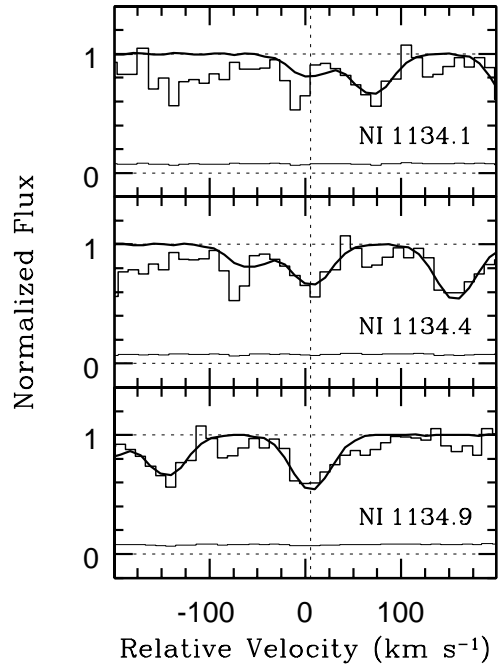}
\caption{Same as Fig.~\ref{FeIIplot}, but for N~{\sc i}. Again, a single-velocity 
component has been modeled with the following parameters: $N$ = 2.8 $\times$ 
10$^{14}$ cm$^{-2}$ and $b$ = 19 km s$^{-1}$ at $v$ = 768 km s$^{-1}$. N~{\sc i} 
$\lambda$1134.1 is the bluest line of the 1134 triplet: it has not been considered 
to constrain the fit parameters due to its unusual shape (artifact/defect of the 
detector?), but it is shown here for completeness.
\label{NIplot}}
\end{figure}

Fe~{\sc ii} is the ion with the best constrained column density determination. Nine absorption lines 
with values of $f\lambda$ spanning a range of 25 from the weakest, $\lambda$1142, to the strongest, 
$\lambda$1144, were simultaneously considered to constrain the fit (Fig.~\ref{FeIIplot}). Moreover, we 
adopted recently updated values for the atomic oscillator strengths $f$ that were empirically determined 
by Howk et al. (2000) in the $FUSE$ band. Fe~{\sc ii} $\lambda$1121 from I~Zw~18 is overlapping with 
Fe~{\sc ii} $\lambda$1125 from the HVC (Fig.~\ref{compspec}), but other Fe~{\sc ii} lines arising from 
the HVC indicate that this contamination is negligible. All the other metals considered in our analysis 
have less stringent constraints. For the ions of the $\alpha$-elements O~{\sc i}, Si~{\sc ii}, and 
Ar~{\sc i}, as well as for N~{\sc i}, a much smaller number of lines was available (Figs.~\ref{alphaplot} 
and ~\ref{NIplot}). In other cases, such as the ions C~{\sc i} and P~{\sc ii}, the absorption was not 
detected, and only an upper limit was inferred for the corresponding column density. Moreover, $f$-values
from the older compilation by Morton (1991) were adopted for the transitions of these ions, the only 
exception being Ar~{\sc i}, for which we considered the newer values from Morton (2003, in preparation).

In the case of O~{\sc i} we used only two lines, $\lambda$976 and $\lambda$1039 (Fig.~\ref{alphaplot}).
All the other O~{\sc i} lines in the $FUSE$ spectral range are contaminated, except for the very weak 
$\lambda$925 that was not considered for the following reasons: {\it i)} the continuum placement is 
highly uncertain due to heavy contamination by other absorptions; {\it ii)} the S/N per resolution 
element is a factor of $\sim$ 2--3 lower than for the other O~{\sc i} lines considered ($\lambda$976 
and $\lambda$1039) due to the fact that $\lambda$925 falls at the edge of the SiC~2A detector where 
the sensitivity is lower; {\it iii)} the oscillator strength has been only theoretically determined 
(see Morton 1991) and never empirically verified, since $\lambda$925 has been rarely observed; moreover, 
some authors report problems associated with this line (e.g., Molaro et al. 2000; Hoopes et al. 2003a) 
and this could be an indication that the theoretical oscillator strength is wrong. We preferred not to 
use O~{\sc i} $\lambda$1302 in the near-UV spectra of comparable resolution from HST/GHRS, since this 
spectrograph has a much smaller aperture sampling only a tiny region of I~Zw~18 and spatial variations 
are expected within the galaxy (e.g., Brown et al. 2002). During the line-profile fitting procedure with
the two selected O~{\sc i} lines ($\lambda$976 and $\lambda$1039) we noticed a degeneracy between column 
density $N$ and Doppler parameter $b$. The weakest O~{\sc i} transition has a value of $f\lambda$ which 
is only 3 times lower than that of the strongest line. The degeneracy for O~{\sc i} is probably due 
to the instrumental resolution of FWHM $\sim$ 35 km s$^{-1}$ coupled with the small range in $f\lambda$ 
values and probable partial saturation. We thus estimated the final column density and Doppler width of 
O~{\sc i} from line-profile fitting with the following procedure. We constrained the interval of possible 
values for both $N$ and $b$ from various acceptable fits, and we chose the center of the corresponding 
interval as our best estimate of a certain parameter and the half width of the same interval as the 
estimated uncertainty. The larger errors associated with the oxygen column density and Doppler broadening 
reflect this procedure and are not related to statistical errors.

Different problems affect the column density determination of Si~{\sc ii}. One of the two lines used 
for the fit, $\lambda$1020, is partially blended with an H$_2$ line, and the continuum in that spectral 
region is slightly affected by the blue wing of Ly~$\beta$ (Fig.~\ref{alphaplot}). The other line, 
$\lambda$989, could be contaminated by N~{\sc iii} $\lambda$989. The absence of N~{\sc ii} $\lambda$1083 
seems, however, to rule out this possibility (although the non-detection of this line could be partly 
due to poor S/N in the corresponding spectral region). On the other hand, the fit obtained by considering 
only Si~{\sc ii} $\lambda$1020 is consistent with the best fit from both absorptions (although with a 
slightly higher $\chi^2$). This could be a further suggestion that contamination of the $\lambda$989 
line by N~{\sc iii} is negligible. In the case of Si~{\sc ii} we do not see the $N$ -- $b$ degeneracy 
problem, even if the strongest $\lambda$989 line is partially saturated, and this is probably due to 
the fact that $f\lambda$ covers a wider range of values (the weakest transition has $f\lambda$ which 
is at least a factor of 4 lower than the strongest line).

Also the column density estimate of Ar~{\sc i} is more uncertain than for Fe~{\sc ii}. The $\lambda$1048 
line is heavily blended with an H$_2$ line, and $\lambda$1066 is very weak (Fig.~\ref{alphaplot}). A 
large error is associated with the Doppler width of the Ar~{\sc i} ion. We have further investigated 
this issue with additional fits, and found out that $b$ can vary over a large range of values, but the
column density is always around the best fit estimate (Table~2). 

Figure~\ref{NIplot} presents the fit we obtained for N~{\sc i}. Only the two reddest lines of the 1134 
triplet were used to this purpose, namely N~{\sc i} $\lambda$1134.4 and N~{\sc i} $\lambda$1134.9 for 
which we have a good constraint. N~{\sc i} $\lambda$1134.1 was not considered for the fit due to its 
unusual shape. This probably arises from a local defect of the LiF~2A detector, whose data we considered 
in this wavelength range. The same feature is in fact weaker and shifted by one pixel ($\sim$ 0.04 \AA) 
in the lower S/N LiF~1B data covering the same spectral range. The two N~{\sc i} lines used are quite weak
and not affected by saturation problems.

The ISM metal lines are found at velocities $v$ = $z\,c$ that are consistent within the errors with 
the systemic velocity of 761 $\pm$ 9 km s$^{-1}$ for the stellar component, after an uncertainty of 
$\delta v$\,$\lesssim$\,6 km s$^{-1}$ ($\delta \lambda$\,$\lesssim$\,0.02 \AA) from the fit centering 
is taken into account. The only exception is the second (reddest) component of Fe~{\sc ii} with a 
velocity of $\sim$ 810 km s$^{-1}$. We want, however, to point out for consistency that the second 
Fe~{\sc ii} component is well within the H~{\sc i} fit of all the Lyman series absorption lines 
considered (Fig.~\ref{lymanplot}). An average value of $v$ = 764 $\pm$ 6 km s$^{-1}$ is obtained 
from the ISM metal lines. This is consistent within the errors with the value of 753 $\pm$ 6 km 
s$^{-1}$ we obtained for H~{\sc i}. We can finally assert that there is no evidence of a velocity 
shift between the absorption lines arising from heavy elements in the neutral ISM of I~Zw~18 and 
its stellar features. This implies that on average there is no evidence of gas outflow/infall over 
the large region of the galaxy sampled by $FUSE$. 

The values of the Doppler width $b$ that fit the ISM metal lines are all comparable, within the 
errors, to the instrumental broadening of 21 $\pm$ 3 km s$^{-1}$. The only exceptions are the 
two velocity components of Fe~{\sc ii} with $b$ $\simeq$ 9 and 3 km s$^{-1}$ ($b$ $\simeq$ 10 
km s$^{-1}$ if only one component is considered for Fe~{\sc ii}). Moreover, $b$ seems
to scale with the ion mass, once errors are correctly taken into account. Indeed the heavier ion 
Fe~{\sc ii} has smaller $b$ values than Si~{\sc ii}, and the latter than the lighter ion N~{\sc i}. 
This trend resembles the case when thermal broadening is higher or comparable to turbulent broadening. 
Large uncertainties instead affect the $b$ estimate of O~{\sc i} and Ar~{\sc i}. If we assume that 
also these ions behave consistently, we should have the Doppler parameter between $\sim$ 16 and 19 
km s$^{-1}$ for O~{\sc i} and between $\sim$ 10 and 16 km s$^{-1}$ for Ar~{\sc i}. The line-profile 
fitting would give a column density of $\sim$ 16.0 dex for O~{\sc i} if a mean value of $b$ $\sim$ 
18 km s$^{-1}$ is adopted, and a column density of $\sim$ 13.6 dex for Ar~{\sc i} if a mean value 
of $b$ $\sim$ 13 km s$^{-1}$ is considered. This is in agreement with what assumed in Table~2 for 
$N_{\rm PF}$. However, we want again to point out that in the one- or two-velocity approximation 
models, $b$ is an ``effective'' Doppler parameter with no real physical meaning (not simply related 
to turbulent and/or thermal broadening), since the observed absorption lines arise from the combination 
of many unresolved components.

In order to investigate the goodness of our single- or double-component approximations, we also performed
the following test. We progressively added one velocity component at a time and fitted the Fe~{\sc ii} 
lines. We were able to reasonably constrain up to 5 velocity components with $b$ in the range 0.6\,$-$\,4 
km s$^{-1}$. We observed, however, the following trend: the total column density is always consistent 
within $1 \sigma$ with the value of 15.09 $\pm$ 0.06 dex of our best two-velocity component fit, even 
if the corresponding error increases with the number of velocity components (i.e., $\log\,N_{\rm PF}$ = 
15.25 $\pm$ 0.14 dex for five components).

We finally derived total column densities by applying the apparent optical depth method (\S~3) in 
order to discover hidden saturation in the unresolved components, as well as to test the one- or 
two-velocity approximation models with a measurement independent of the velocity structure of the 
absorbers. Column densities of I~Zw~18 absorption lines estimated with this method are listed as 
$N_{\rm AOD}$ in column 9 of Table~2, the only exception being Si~{\sc ii} $\lambda$989, which is 
saturated (if $I_0(v)$ approaches 0, $\tau_{a}(v)$ becomes undetermined and the method is no longer 
applicable). Column 8 gives the velocity interval $\Delta v$ over which equation (3) was calculated. 
By inspecting the values of $N_{\rm AOD}$ for Fe~{\sc ii}, it is evident that the two strongest 
lines, $\lambda$1144 and $\lambda$1063, have a much lower apparent column density indicative of 
possible saturation. On the other hand, the average over the remaining transitions gives a value 
for $N_{\rm AOD}$ of 14.99 $\pm$ 0.14 dex, which is consistent within the errors with the value of 
$N_{\rm PF}$. For O~{\sc i}, the lower value of $N_{\rm AOD}$ in the line $\lambda$1039 suggests 
saturation problems. Nevertheless, the other line, $\lambda$976, gives a value for $N_{\rm AOD}$ 
of 15.96 $\pm$ 0.12 dex, which is in agreement with profile fitting measurements. The only unsaturated 
Si~{\sc ii} line, $\lambda$1020, is blended and the corresponding $N_{\rm AOD}$ is a lower limit 
consistent with our profile fitting measurements. Two lines are available for Ar~{\sc i}. The blended 
$\lambda$1048 line gives a lower limit for the column density, while the weak $\lambda$1066 line has 
a value of $N_{\rm AOD}$ in agreement with $N_{\rm PF}$. N~{\sc i} has two absorptions that we used 
to infer the column density, and the mean value for $N_{\rm AOD}$ of 14.46 $\pm$ 0.05 dex is again 
consistent with the profile fitting results. 

The application of the apparent optical depth method strengthens our assumption that in general 
the velocity models that we considered for the line-profile fitting are good representations of 
the $FUSE$ data, even if the real physical situation is more complex. In the following, we will 
thus adopt $N_{\rm PF}$ for the final column densities of all ions in the neutral ISM of I~Zw~18. 
However, we caution the reader that O~{\sc i} could be an exception. It is important to notice 
that the strengths of the O~{\sc i} lines differ by a factor of 3, yet there is still a 0.35 dex 
difference in $N_{\rm AOD}$. The true O~{\sc i} column density may thus be underestimated by as 
much as 0.5 dex, if the apparent column densities listed in Table~2 are corrected for unresolved 
saturation as recommended by Savage \& Sembach (1991). This would happen if the velocity distribution 
of the neutral gas contains very strong components that are substantially under-resolved by $FUSE$. 
Such an upward correction of the O~{\sc i} column density is still consistent within $2 \sigma$ with 
the profile fitting results and would bring the O~{\sc i} abundance in the neutral gas up to a value 
closer to that in the H~{\sc ii} regions. However, in the absence of additional information we will 
assume that the line-profile fitting and the apparent optical depth results for the weak line of 
O~{\sc i} ($\lambda976$) give the correct column density for O~{\sc i}.

\subsection{Abundance Determinations and Correction Effects}

The major concern in abundance determinations from UV absorption-line analysis is represented 
by ionization and dust-depletion correction effects. 

Ionization effects are usually neglected, and abundances are derived by assuming that the primary 
ionization state of an element in the neutral gas is representative of the total amount of the 
element. From Galactic interstellar studies it is well known that the singly ionized stage is the 
dominant one for most elements because their first ionization potential is below 13.6 eV (the H$^0$ 
ionization threshold) and their second one is above it. The neutral stage instead prevails for those 
elements having the first ionization potential above 13.6 eV. The reason for this is that the bulk 
of the H~{\sc i} gas with $N_{\rm HI}$ $\gtrsim$ 10$^{19}$ cm$^{-2}$ is self-shielded from $h\nu$ 
$>$ 13.6 eV\,photons, but transparent to $h\nu$ $<$ 13.6 eV\,photons. This means that Fe~{\sc ii}, 
C~{\sc ii}, Si~{\sc ii}, and P~{\sc ii}, as well as O~{\sc i}, N~{\sc i}, and Ar~{\sc i} will be 
the dominant ionization stages of these elements in the neutral gas of I~Zw~18.

Some of the ions that are dominant ionization states in H~{\sc i} regions may also be produced in 
photoionized clouds where H~{\sc i} is a small fraction of the total hydrogen content. The formation 
of metal absorption lines in both ionized and neutral regions can have a significant impact on element 
abundance determinations. This problem has been recently tackled in detail for damped Ly$\alpha$ systems 
(DLAs) by Howk and Sembach (1999) and Vladilo et al. (2001) using the CLOUDY code (Ferland et al. 1998). 
These authors confirm previous findings that corrections to interstellar abundances are negligible for 
the majority of elements observed in DLAs in the case of high column densities, i.e. N$_{\rm HI}$ $\gtrsim$ 
10$^{21}$ cm$^{-2}$ (but see Prochaska et al. 2002, or Izotov et al. 2001 for a contrasting opinion on 
this issue). The ionization corrections are in general smaller when a stellar spectrum dominates over 
an external UV background as ionizing source of the ISM. This is the case in I~Zw~18, where the radiation 
field is produced by its young massive stars.

\begin{figure}
\plotone{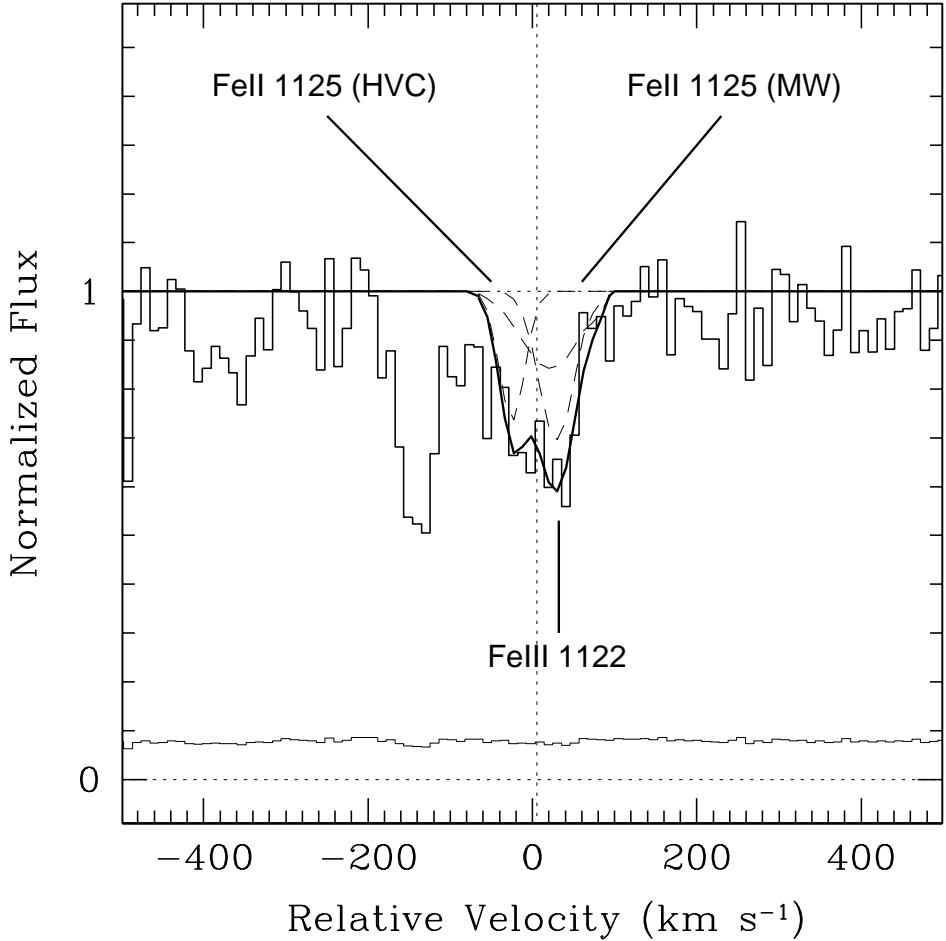}
\caption{Fit of the Fe~{\sc iii} absorption line arising from I~Zw~18: observed spectrum 
({\it histogram}) and theoretical line profiles ({\it solid and dashed lines}). Fe~{\sc iii}
(deep dashed line on the right) is blended with Fe~{\sc ii} $\lambda$1125 from the MW (shallow 
dashed line on the right) and the HVC (deep dashed line on the left). The solid line represents
the composite fit. The contaminating absorption features were well constrained by simultaneous 
line-profile fitting of other uncontaminated Fe~{\sc ii} lines from, respectively, the MW and 
the HVC at different wavelengths. A final column density of $N$ = 4.0 $\times$ 10$^{13}$ cm$^{-2}$ 
was derived for the Fe~{\sc iii} line. The spectrum of the noise per pixel is also shown ({\it 
thin line histogram}).
\label{FeIIIplot}}
\end{figure}

The relative mixture of neutral and ionized gas contributing to an absorbing spectrum can be 
empirically determined by measuring adjacent ions of the same elements, e.g., Fe~{\sc ii}/Fe~{\sc iii} 
or Al~{\sc ii}/Al~{\sc iii} (Howk \& Sembach 1999; Sembach et al. 2000; Prochaska et al. 2002). In the 
case of I~Zw~18 we observed Fe~{\sc iii} $\lambda$1122. Fe~{\sc iii} absorption is purely interstellar 
when a hot young O stellar population dominates the $FUSE$ spectral range (Walborn et al. 2002; Pellerin 
et al. 2002), and only originates in the ionized gas associated with H~{\sc ii} regions. The regions with 
neutral gas are instead where the bulk of Fe~{\sc ii} absorption is produced. We followed the recipe 
of Sembach et al. (2000) to correct our derived Fe~{\sc ii} abundances for ionization effects. Since 
the $FUSE$ aperture integrates over a large area, we considered their ``composite'' model for the warm 
ionized ISM of the MW (a combination of overlapping low-excitation H~{\sc ii} regions) as a better 
representation of the ionized gas conditions in I~Zw~18. Unfortunately, Fe~{\sc iii} $\lambda$1122 is 
blended with Fe~{\sc ii} $\lambda$1125 from the MW and the HVC, but we could take this into account with 
the line-profile fitting. The total column density of Fe~{\sc iii} we inferred from our measurements is 
13.61 $\pm$ 0.08 dex (Fig.~\ref{FeIIIplot}). This is also the Fe~{\sc ii} contribution from the ionized 
gas, since half of the iron is in Fe~{\sc iii} and half in Fe~{\sc ii} (Table~5 of Sembach et al. 2000). 
This implies that $\lesssim$\,13.70 dex of the total Fe~{\sc ii} column density (15.09 $\pm$ 0.06 dex) 
measured in the $FUSE$ spectra might be due to ionized gas, corresponding to an ionization fraction of 
at most 5\%. If we correct for this factor, we get an Fe~{\sc ii} column density in the neutral gas of 
15.07 dex, just 0.02 dex lower than measured, and well within the 1$\sigma$ errors. The radiation field 
is likely to be much higher in I~Zw~18 than in the MW thanks to the higher density of younger and hotter 
stars responsible for it. In addition, the ISM metallicity in I~Zw~18 is much lower than in the MW, 
this property implying higher temperatures for the gas since less heavy elements are available for 
cooling. Both effects go in the direction of favoring more iron in the Fe~{\sc iii} ionization state 
than in the Fe~{\sc ii}, thus much lower ionization corrections. For example, if we consider for our 
model a higher ionization parameter of $\log (q)$ = $-2$ instead of the value of $-4$ of the MW, we 
obtain that $\sim$ 70\% (and not 50\%) of the iron is Fe~{\sc iii}, corresponding to an ionization 
fraction less than 2\%. Fe~{\sc iii} is thus a very good tracer of ionized gas (much better than 
Al~{\sc iii}), since it is the dominant state of iron for a wide range of ionizing conditions (e.g., 
effective temperature of the ionizing central star or ionization parameter). 

The ionization correction for Fe~{\sc ii} is one of the largest: in the ionized gas all the other elements 
like H, O, N, and Ar are found mostly ($\gtrsim$ 80\%) in an ionization state higher ({\sc ii}) than the 
one we detect in the UV spectra ({\sc i}). We can thus assess that ionization effects should be negligible 
in our $FUSE$  data. The only exception is Si, where as much as $\sim$ 90\% of Si~{\sc ii} (70\% for a more 
plausible higher ionization parameter of $\log (q)$ = $-2$) could arise in the ionized gas. However, no 
absorption lines of Si~{\sc iii} are available in the $FUSE$ band to correctly take this into account. 

The presence of dust in the ISM can represent another serious complication in the interpretation of 
the metal abundances. Refractory elements (e.g., Fe and Si) are more easily locked into dust grains 
than non-refractory ones (e.g., O, Ar, N). This can clearly alter the total heavy element abundances. 
Observations of the local ISM provide hints of selective depletions acting in dense clouds within our 
Galaxy (Savage \& Sembach 1996). There is evidence of dust also in the DLAs, systems that I~Zw~18 
resembles in its H~{\sc i} column density. The reddening of QSOs lying behind DLA absorption is a clear 
clue of dust extinction (Pei, Fall, \& Bechtold 1991; Fall \& Pei 1993). Evidence of elemental depletion 
similar to that observed in the nearby ISM was first reported for DLAs by Pettini et al. (1994), and 
further supported by others (Hou, Boissier, \& Prantzos 2001; Prochaska \& Wolfe 2002), although with 
a much lower dust-to-gas ratio (typically $\sim$ 1/30) than in our own Galaxy (Pettini et al. 2000). 
Molecular hydrogen has also been found recently in 15\%-20\% of a sample of DLAs and usually in those 
systems with the highest metallicities and largest dust depletions (Ledoux, Petitjean, \& Srianand 
2003). On the contrary, I~Zw~18 does not show detectable $H_2$.

A detailed study of the dust content of I~Zw~18 from its H~{\sc ii} regions has been performed recently 
by Cannon et al. (2002). A bidimensional map of the extinction has been obtained from the ratio between 
H$\alpha$ and H$\beta$ HST/WFPC2 images. The optical extinction, $A_{\rm V}$, varies significantly from 
one region to another, suggesting a patchy dust distribution, but the extinction is always small, with 
a maximum value of $\sim$ 0.5 mag and a mean value of only $\sim$ 0.15 mag. This is not an unexpected 
trend, since low-metallicity galaxies should have a lower and more patchy dust content (Morgan \& Edmunds 
2003).

Dust traced by emission-line spectra does not necessarily correspond to dust content in the neutral 
ISM (different geometries and different regions are involved). This is the reason why it is important
to have an indication of depletion directly from our $FUSE$ absorption measurements. The same methods
applied to DLAs could be used for this purpose. In general, dust properties are often assumed to be 
similar to those of the Galactic interstellar dust, and the amount of depletion is simply scaled by 
taking into account variations of the dust-to-metal ratios (Vladilo 1998; Savaglio 2001). However, the 
observed abundances arise from the combination of {\it i)} a non-solar nucleosynthesis, {\it ii)} a 
non-solar star-formation history, and {\it iii)} a certain dust depletion pattern. In order to disentangle 
these effects, refractory and non-refractory elements originating from similar nucleosynthetic histories 
should be considered in the analysis, i.e., the Fe-peak group (e.g., Cr, Zn and Fe). Unfortunately, only 
the Fe abundance has been measured in the neutral ISM of I~Zw~18.

In the absence of a possible measurement of the dust content in the neutral ISM of I~Zw~18, we will 
not consider depletion corrections in our following analysis. This is clearly an approximation, but 
we do not expect significant corrections due to the very low metal content of the galaxy. Moreover,
among all the chemical elements considered, only Si and Fe are sensitive to this problem. For reference, 
in DLAs the depletion corrections can be as high as $\sim$ 0.2$-$0.4 dex for Si, and $\sim$ 0.7$-$1.1 
dex for Fe (Garnett 2002). We note that we can use oxygen and argon (neither of which will be depleted)
to determine the abundances of the $\alpha$ elements.

In Table~3 we summarized the column densities of ions representative of the total amount of an element 
in the neutral phase of the ISM (column 3). Only an upper limit is available for P~{\sc ii}. C~{\sc i} 
($\log N$ $<$ 14.40 at $3 \sigma$ level) was omitted, since C~{\sc ii} is the dominant state for carbon. 
However, the only C~{\sc ii} absorption line 
in the $FUSE$ spectral range, $\lambda$1036, is heavily contaminated by O~{\sc i} $\lambda$1039 from 
the MW. This implies that we were unable to determine the carbon abundance in the neutral ISM of I~Zw~18. 
Column densities were converted into abundances relative to $N$(H) by adopting the standard solar 
photospheric abundances published by Grevesse, Noels, \& Sauval (1996), and the formula [X/H] = $\log$(X/H) 
$-$ $\log$(X/H)$_{\odot}$ (column 6). The only exception is oxygen, for which an updated solar value of 
12\,+\,$\log$(O/H) = 8.69 $\pm$ 0.05 dex ($\sim$ 0.2 dex lower than previously believed) from Allende Prieto, 
Lambert, \& Asplund (2001) was used. Errors on [X/H] reflect measurement errors in $N$(X). For a direct 
comparison, column~7 and 8 of Table~3 present abundances for the ionized gas in the northwest (NW) and 
southeast (SE) H~{\sc ii} regions of I~Zw~18. These have been homogenized to the neutral gas abundances 
by rescaling the values from Izotov et al. (1999) and Izotov and Thuan (1999) with our adopted solar scale. 
In this way, the oxygen abundance of the H~{\sc ii} regions turns out to be $\sim$ 1/33 solar compared to 
the old value of $\sim$ 1/50.

\section{Discussion}

\subsection{Comparison with Abundances in the H~{\sc ii} Regions} 

Our $FUSE$ observations sample absorption lines of three $\alpha$-elements, O, Ar and Si. Oxygen 
in the H~{\sc ii} regions is the most secure abundance measurement of all the elements whose ions 
have been detected in the different ISM phases of I~Zw~18. On the other hand, oxygen in the neutral 
gas phase has the highest uncertainty, and could be affected by systematic errors that could increase 
its abundance by as much as $\sim$ 0.5 dex (\S~7.1). However, if we take our estimate at face value, 
we infer for the neutral ISM an [O/H] = $-$\,2.06 $\pm$ 0.28 dex, which is $\sim$ 3--4 times lower 
than in the ionized gas (Table~3). A similar trend is found for oxygen in the BCD Mrk~59 by Thuan 
et al. (2002). The difference between H~{\sc i} and H~{\sc ii} regions (if not related to systematic 
uncertainties) cannot be explained with ionization effects in the neutral gas, since O~{\sc i} is 
tightly related to H~{\sc i}, so that [O~{\sc i}/H~{\sc i}] = [O/H]. It cannot be explained with 
dust depletion either, since O is non-refractory, and it can only be the result of nucleosynthesis.

The argon abundance in the H~{\sc ii} regions of I~Zw~18 is on average [Ar/H] = $-$\,1.51 $\pm$ 0.08 
dex, a factor of $\sim$ 6 times higher than the value measured in the neutral ISM (Table~3). Part of 
the difference between neutral and ionized gas could be due to ionization effects. Ar does not 
generally behave like H (and O) in the ISM despite similar ionization potentials (Sofia \& Jenkins 
1998; Jenkins et al. 2000). Its larger photoionization cross section renders much easier for Ar 
to hide in its ionized form, Ar~{\sc ii}, if photons with $h\nu$ $>$ 15.76 eV (the Ar~{\sc i} 
ionization threshold) are able to leak through a H~{\sc i} layer (i.e., low column density regions
with $N$(H~{\sc i}) $\lesssim 10^{19}$ cm$^{-2}$). As a result the [Ar~{\sc i}/H~{\sc i}] ratio is 
lower than the real [Ar/H]. However, due to the relatively high column densities of the ISM, we 
expect ionization effects to be negligible in I~Zw~18. If ionization 
is not important, we should observe the same value of $\log$\,(Ar/O) in the different gas phases of
I~Zw~18, since both Ar and O are non-refractory and are produced by the same stars. And indeed the 
value of $\log$\,(Ar/O) = $-$\,2.16 $\pm$ 0.08 dex for the ionized gas seems to be preserved within 
the errors in the neutral phase, where $\log$\,(Ar/O) = $-$\,2.38 $\pm$ 0.27 dex, in agreement with 
the predictions of standard stellar nucleosynthesis models (e.g., Woosley \& Weaver 1995). The 
$\log$\,(Ar/O) ratio of I~Zw~18 is well within the range ($-$\,2.00 to $-$\,2.5) found in both low 
(van Zee, Haynes, \& Salzer 1997) and high (Thuan, Izotov, \& Lipovetsky 1995) surface brightness 
star-forming dwarf galaxies. It is also interesting that the Ar/O ratio in the neutral gas of I~Zw~18 
(as well as in its H~{\sc ii} regions) has practically a solar value ([Ar/O] = $-$\,0.21 $\pm$ 0.31 
dex), and is very similar to the interstellar value of [Ar/$\alpha$] $\sim$ $-$\,0.2 dex estimated 
by Vladilo et al. (2003) for the local universe.

The other $\alpha$-element, silicon, behaves in a slightly different manner. Its abundance of [Si/H] 
= $-$\,2.09 $\pm$ 0.12 dex in the neutral ISM is only marginally lower that in the H~{\sc ii} regions 
(Table~3). Depletion cannot explain the lack of a clear Si excess in the H~{\sc ii} regions (even 
if it can justify a lower Si content compared to the other $\alpha$-elements). Dust grains are more 
easily destroyed (than formed) by star-formation through shocks generated by supernova explosions 
and propagating into the ISM (Savage \& Sembach 1996). The following possibilities must instead be 
considered: {\it i)} our abundance is overestimated due to contamination problems affecting the 
Si~{\sc ii} lines (\S~7.1); {\it ii)} ionization corrections are not negligible for this element, 
at variance with what we assumed (\S~7.2), and the real abundance in the neutral gas is lower than 
inferred; {\it iii)} the abundances in the H~{\sc ii} regions are higher than reported (note the 
large uncertainties given in Table~3); {\it iv)} part of the Si in the neutral gas has been produced 
by SNe~Ia, not just SNe~II, even if this contribution should be small. On the other hand, we cannot 
completely rule out that the different ISM phases are more homogeneous than indicated by the other 
$\alpha$-elements, O and Ar.

The iron abundance in the neutral ISM of I~Zw~18 is one of our most reliable estimates. On the other 
hand, the Fe content in the southeast H~{\sc ii} region (Izotov et al. 1999) is likely more uncertain 
than suggested by the formal error of Table~3 (0.09 dex). The iron abundance is derived from a single 
Fe$\,^{2+}$ emission line ([Fe~{\sc iii}] $\lambda4658$), but most of the iron is in Fe$\,^{3+}$. Thus,
an ionization correction factor (ICF) has to be applied in order to infer the total amount of this 
element from its minority species. It is possible that the ICF value is more uncertain than 
quoted (the authors do not formally report any error). This is suggested, e.g., by the scatter of 
the ICF for different regions of the other BCD analyzed in the same article, SBS~0335-052. 
The Fe content in the neutral ISM of I~Zw~18 is, however, consistent, within the uncertainties, with 
that in the ionized gas. Iron, thus, behaves differently from the $\alpha$-elements O and Ar, whose 
abundances are several times lower in the neutral gas, and more similarly to the third $\alpha$-element 
considered, Si.

The last element directly comparable in H~{\sc i} and H~{\sc ii} regions is nitrogen. The value of [N/H] =
$-$\,2.88 $\pm$ 0.11 we derived for the neutral gas in I~Zw~18 is a factor of $\sim$ 3 lower than in the 
southeast H~{\sc ii} region. Ionization (but not depletion) effects could be responsible for part of this 
difference, since for $N$(H~{\sc i}) $\lesssim 10^{19}$ cm$^{-2}$ N~{\sc i} is not strongly coupled to 
H~{\sc i} and behaves more similarly to Ar~{\sc i}, even if to a smaller extent (Sofia \& Jenkins 1998;
Jenkins et al. 2000). However, in the nearby starburst galaxy NGC~1705, where  $N$(H~{\sc i}) $\simeq 
10^{20}$ cm$^{-2}$, Heckman et al. (2001) were able to directly address this issue with detections of 
both N~{\sc i} and N~{\sc ii}, and indeed they found that most of the observed N~{\sc ii} must be 
associated with the warm photoionized gas. In our case, we do not detect N~{\sc ii} $\lambda$1083 
(\S~7.1), so we expect ionization effects to be negligible. In addition, we also note that the solar 
value of Ar/O in both the neutral and ionized ISM of I~Zw~18 rules out ionization effects for Ar in 
the H~{\sc i} region, and this is an even stronger result for N. We thus conclude that the N 
underabundance in the neutral ISM of I~Zw~18, compared to its H~{\sc ii} regions, is real and only 
due to stellar nucleosynthesis.

\subsection{The Chemical Evolutionary History of I~Zw~18}

What can we infer about the evolutionary state of I~Zw~18 from the metal content of its interstellar
gas? The first straightforward conclusion is that the abundances of heavy elements in its neutral ISM
are low, but not zero: $\alpha$-elements are between $\sim$ 1/100 and 1/200 solar, Fe is $\sim$ 1/60 
solar, and N is $\sim$ 1/800 solar. According to Panagia (2002), a metallicity above 10$^{-3}$ 
Z$_{\odot}$ for elements arising from SNe~II (i.e., $\alpha$-elements) is the result of one or more 
episodes of metal enrichment. This implies that the neutral gas in I~Zw~18 has already been enriched 
with the products of at least one star-formation event, and is not primordial in nature. However, its 
extremely low abundances do not leave much room for a large amount of star-formation. 

The abundance pattern we observe in the different gaseous components of I~Zw~18 directly leads to a 
second conclusion. The metal content is significantly lower in the H~{\sc i} than in the H~{\sc ii} 
regions, apart from iron which is the same. The $\alpha$-elements are mainly produced by SNe~II on 
short time scales (less than 50 Myr), while iron is mostly released by SNe~Ia on time scales longer 
than 1 Gyr. This may indicate that the H~{\sc ii} regions, compared to the neutral ISM, have been 
additionally enriched in $\alpha$-elements (and nitrogen) by more recent star formation activity.

Can we invoke self-enrichment of the H~{\sc ii} regions by their massive stars? In general, the concept of 
self-enrichment (Kunth \& Sargent 1986) does not agree with the homogeneity observed in H~{\sc ii} regions 
in dwarf star-forming galaxies (e.g., Kobulnicky \& Skillman 1997). I~Zw~18 is not an exception from this 
point of view (Skillman \& Kennicutt 1993; V\'{\i}lchez \& Iglesias-Par\'amo 1998; Legrand et al. 2000). In 
addition, Larsen, Sommer-Larsen, \& Pagel (2001) present theoretical evidence that the supershells triggered 
by SNe~II and stellar winds from massive stars are always well within the H~{\sc ii} regions for almost the 
whole existence of these regions (less than 10 Myr). This implies that elements newly synthesized in an ongoing 
burst are trapped in the hot phase of these superbubbles and can never be well mixed with the emitting gas 
within the H~II region. It is thus highly unlikely that we are witnessing direct self-enrichment in the H~II 
regions of I~Zw~18. 

A key element for shedding light on the chemical evolution of I~Zw~18 is nitrogen. The nature and origin 
of nitrogen in galaxies has been debated for a long time. In massive metal-rich galaxies, N/O increases 
roughly linearly with O/H (Torres-Peimbert, Peimbert, \& Fierro 1989; van Zee, Salzer, \& Haynes 1998a). 
This is the typical behavior for ``secondary'' nitrogen, produced by intermediate-mass stars on time scales 
longer than $\sim$ 250 $-$ 300 Myr. In contrast, dwarf star-forming galaxies (Thuan et al. 
1995; Kobulnicky \& Skillman 1996; van Zee et al. 1997) and low-metallicity H~{\sc ii} regions 
in spirals (van Zee et al. 1998a) show a relatively constant N/O {\it versus} O/H with the plateau 
around $\log$\,(N/O) = $-$\,1.5. This is attributed to ``primary'' nitrogen, mainly produced by 
intermediate-mass stars (Renzini \& Voli 1981; Meynet \& Maeder 2002a), but also by low-metallicity massive 
($M\,>\,$30 M$_{\odot}$) stars (Woosley \& Weaver 1995;  Meynet \& Maeder 2002b) on much shorter time scales 
($\sim$ 10 Myr). The nitrogen abundance in I~Zw~18 is higher in the H~{\sc ii} regions than in the neutral 
ISM, consistent with what is found for $\alpha$-elements. This indicates that there has been metal enrichment 
by star formation that occurred long enough ago for primary nitrogen to have been released back into the ISM. 
This dates the star formation to at least $\sim$ 10 Myr ago (if massive stars are responsible), or several 
hundred Myr ago (if intermediate-mass stars are responsible). The simplest interpretation would be that the 
current episode of star-formation traced by the H~{\sc ii} regions has been going on long enough for nitrogen 
enrichment to have occurred in this part of the galaxy ISM.

We can place time constraints on the different star-formation episodes with simple considerations of the 
physical scales involved. The nucleosynthetic activity responsible for the enrichment in the H~{\sc ii} 
regions must have taken place long enough ago for gas mixing to be efficient in the central optical part 
of the galaxy (as testified by the homogeneity of the H~{\sc ii} regions; Izotov et al. 1999). The linear 
scales involved are around $\sim$\,0.5 kpc, corresponding to mixing time scales of $\sim$ 100 Myr (Roy \& 
Kunth 1995). It must have also started recently enough so that mixing on physical scales comparable to the 
extended H~{\sc i} envelope surrounding the star-forming regions is incomplete. The halo of I~Zw~18 is 
$\sim$\,5 kpc wide from H~{\sc i} maps of van Zee et al. (1998b). This corresponds to mixing time scales 
around $\sim$ 1 Gyr (Tenorio-Tagle 1996). We thus conclude that the star-formation process responsible for 
the additional enrichment of the H~{\sc ii} regions must have started more recently than 1 Gyr ago and have 
been active for at least a hundred Myr. This rules out contribution to primary nitrogen only from massive 
stars with very short lifetimes ($\sim$ 10 Myr). By the same reasoning, the ancient star-formation episode 
responsible for the metals in the H~{\sc i} must have started at least $\sim$ 1 Gyr ago to allow for 
diffusion of metals from the central stellar production sites.

The times of the various star-formation activities can be further constrained by considering the behavior of 
abundance ratios involving elements produced on different time scales. One of the most powerful tools from this 
point of view is O/Fe. According to Gilmore \& Wyse (1991), at the beginning of the very first star-formation 
activity we should find a value of [O/Fe]\,$\simeq$\,+\,0.63 dex (+\,0.45 on the old oxygen solar scale) which 
reflects the relative production of O and Fe from a mean SN~II. As soon as the first SNe~Ia start to explode, 
this ratio begins to decrease due to additional Fe contributions. The decrement is more rapid in the case of 
a short burst than for continuous star-formation. The onset of another burst will simultaneously, and almost 
instantaneously, increase the contribution to the ISM of both O and Fe from SNe~II. The O/Fe ratio will suddenly 
jump to a higher value, even if always lower than the initial value due to the continuous contribution from SNe~Ia 
to Fe only. In the neutral ISM of I~Zw~18 we infer a value of [O/Fe] = $-$\,0.30 $\pm$ 0.30. This clearly indicates 
a substantial Fe enrichment from SNe~Ia in the H~{\sc i} gas, and ages longer than 1 Gyr for the oldest episode. 
On the other hand, a supersolar value of [O/Fe]\,=\,+\,0.45 $\pm$ 0.09 is found in the southeast H~{\sc ii} 
region, where the Fe/H ratio (a metallicity indicator) is instead the same within the errors. This implies that 
more recent and relatively young star-formation activity (with no contribution from additional SNe~Ia, thus with 
an age less than 1 Gyr) is responsible for the further enrichment of the ionized gas. The epochs we infer by 
analyzing the behavior of O/Fe are thus in agreement with our considerations of the mixing processes for both 
the ancient and more recent star-formation episodes.

The O/Fe results on the age of the ancient star-formation activity are quite strong. They are still valid 
even if iron is partly depleted onto dust grains, since the contribution from SNe~Ia would be higher in this 
case. Also infall of primordial gas (e.g., Matteucci \& Chiosi 1983; Edmunds 1990) would not alter this 
conclusion, as its net effect would be the dilution of the whole metal content and would not affect relative 
abundances. What could influence these findings is the presence of gas outflow, especially if it is selectively 
enriched in $\alpha$-elements from SNe~II (Pilyugin 1993; Marconi, Matteucci, \& Tosi 1994; Bradamante, 
Matteucci, \& D'Ercole 1998; Recchi et al. 2002; but see also Recchi, Matteucci, \& D'Ercole 2001 for an 
alternative enrichment in iron from SNe~Ia). The loss of oxygen through galactic winds would decrease the 
O/Fe ratio in a way that could mimic the iron contribution from SNe~Ia. However, our $FUSE$ data provide no 
observational evidence of a gas outflow in I~Zw~18. Martin (1996) discusses the existence of a bipolar H$\alpha$ 
bubble with a dynamical age of 15$-$30 Myr that will eventually evolve into a galactic wind. This does not imply 
that part of the ISM has already been ejected from the galaxy. From their analysis of the O~{\sc vi} 
$\lambda\lambda$1032, 1038 doublet, Heckman et al. (2002) find no indication of a ``warm'' coronal phase 
associated with I~Zw~18, although this could reflect the very low oxygen abundance. The lack of large-scale 
outflows in the early stages of I~Zw~18 evolution is theoretically predicted by Recchi et al. (2002). In 
their chemical evolution model, a galactic wind enriched in SNe~II products sets in only in the last 15 
Myr with the second and more intense of two bursts of star formation. 

Under the assumption of no gas outflow, we calculated the number of SNe~Ia that have enriched 
the neutral ISM of I~Zw~18. The following procedure was applied. The Fe abundance in the H~{\sc i} gaseous 
component probably arises from both SNe~II and SNe~Ia. We disentangled their relative contribution by taking 
into account the quantity of oxygen produced by SNe~II during the star-formation event and the prescriptions 
by Gibson, Loewenstein, \& Mushotzky (1997). This amounts to $\sim$\,50\% for SNe~Ia and $\sim$\,50\% for 
SNe~II in the case of [O/Fe] = $-$\,0.30 ($-$\,0.12 on the old oxygen solar scale). We obtained a total Fe 
mass of $\sim$ 700 M$_{\odot}$ by assuming a total H~{\sc i} mass of $\sim$ 2.6 $\times$ 10$^7$ M$_{\odot}$ 
(van Zee et al. 1998b) and an Fe abundance equal to $\sim$ 1/60 solar ([Fe/H] = $-$\,1.76). Only $\sim$\,50\% 
of this mass has been produced by SNe~Ia. Considering a typical SN~Ia yield of $\sim$ 0.74 M$_{\odot}$ for 
iron (Gibson et al. 1997), we inferred that $\sim$ 470 SNe~Ia are required in order to justify the metal 
enrichment in the neutral ISM of I~Zw~18.

The N/O ratio can also help us in further investigating the nature and properties of the star-formation 
activity in I~Zw~18. Unlike the Fe ratio, N/O is free of dust depletion. Moreover, in the case of I~Zw~18 
ionization problems do not affect its value since N~{\sc i} should behave like O~{\sc i} (see \S~8.1). We 
have already mentioned the existence of a plateau in the N/O {\it versus} O/H relation for metal-poor 
star-forming galaxies and its explanation as being due to primary nitrogen. Many authors (Edmunds \& Pagel 
1978; Garnett 1990; Larsen et al. 2001; Contini et al. 2002; Chiappini, Romano, \& Matteucci 2003) are able 
to reproduce this plateau with their chemical evolution models by simply considering a bursting regime 
coupled to the time delay between the delivery into the ISM of O from massive stars and N from 
intermediate-mass stars. Primary nitrogen from massive stars is thus not a necessary condition, at 
variance with the early suggestions by Izotov and Thuan (1999). On the other hand, Henry, Edmunds, \& 
K\"oppen (2000) propose a more continuous low-level ($\sim$ 10$^{-3}$ M$_{\odot}$ yr$^{-1}$ kpc$^{-2}$) 
of star-formation activity to justify the existence of the plateau. This last model is more consistent 
with the picture that BCDs have an underlying old (several Gyr) metal-poor stellar population with bursts
superimposed on it (e.g., the chemical evolution model of I~Zw~18 by Legrand 2000). Every galaxy that 
evolves slowly will thus maintain a relatively low metallicity over a significant fraction of a Hubble 
time since the total metallicity is directly related to the star-formation rate integrated over time. 
Values of $\log$\,(N/O) well below the plateau, as observed in some DLAs (Lu, Sargent, \& Barlow 1998; 
Centuri\'on et al. 2003), can be found only in the very early stages of the N/O evolution, when the 
stellar objects have ages younger than or comparable to the typical time scales for the release of 
primary N from intermediate-mass stars (Henry et al. 2000; Centuri\'on et al. 2003).

From our $FUSE$ observations we obtain a value of $\log$\,(N/O) = $-$\,1.54 $\pm$ 0.26 for $\log$\,(O/H) = 
$-$\,5.37 $\pm$ 0.28. The neutral ISM of I~Zw~18 thus follows the flat trend in N/O {\it versus} O/H observed 
for dwarf star-forming galaxies (e.g., Garnett 2002), as well as metal-rich DLAs (Centuri\'on et al. 2003). 
However, it is well above the values found at similar metallicities in those O-poor DLAs considered to be 
chemically young systems (Lu et al. 1998; Centuri\'on et al. 2003). Hence, H~{\sc i} gas in I~Zw~18 seems 
to have been polluted by star-formation at least as old as the typical lag time (250$-$300 Myr) for N released 
by intermediate-mass stars, and probably older than 1 Gyr (based on the results for O/Fe). A value of 
$\log$\,(N/O) = $-$\,1.57 $\pm$ 0.06 for $\log$\,(O/H) = $-$ 4.82 $\pm$ 0.03 is inferred for the southeast 
H~{\sc ii} region of I~Zw~18. Such a similar N/O but a higher O content in the ionized gas compared to the 
neutral ISM are possible only if the galaxy is experiencing or has experienced additional star-formation 
which has released additional oxygen from SNe~II but also additional nitrogen from massive or 
intermediate-mass stars. 

Silicon is the last element that we can consider to constrain the star-formation history of I~Zw~18. It 
is an $\alpha$-element produced mainly by SNe~II, but behaves differently from O and Ar (see \S~8.1). For 
example, Si is depleted onto dust grains in studies of the local ISM, while Ar and O are non-refractory 
(Savage \& Sembach 1996). Izotov \& Thuan (1999) have, however, found that Si/O is constant with O/H in 
their sample of BCDs, and claimed that the similarity of this plateau with the solar value is evidence 
of no depletion in the H~{\sc ii} regions. Silicon is also more affected by ionization corrections than 
O and Ar, since Si~{\sc ii} is the majority species in intervening H~{\sc ii} gas as well. Moreover, a 
small amount of silicon is produced by SNe~Ia (Matteucci, Molaro, \& Vladilo 1997), at variance with the 
other $\alpha$-elements considered in our analysis. The silicon abundance in the neutral gas of I~Zw~18 
is similar to the oxygen one. On the other hand, Si in the H~{\sc ii} regions is lower than the other 
$\alpha$-elements. Two interpretations are possible to explain these observations: {\it i)} the Si 
value in the ionized gas is higher than estimated; {\it ii)} Si is depleted in the H~{\sc ii} regions, 
and as a consequence in the neutral gas as well; however, no depletion is noticed in the neutral ISM 
because ionization is important for Si~{\sc ii} or some contribution from SNe~Ia is present. In general, 
silicon tracks very well the other $\alpha$-elements O and S, so that O/Si $\simeq$ S/Si $\simeq$ 0
(indicative of the same SNe~II origin). This is, e.g., what is observed in dwarf star-forming galaxies 
(Izotov \& Thuan 1999) and DLAs (Prochaska \& Wolfe 1999; Vladilo et al. 2003). Thus, the first 
interpretation is probably the one to prefer since [O/Si] = +\,0.03 $\pm$ 0.30 in the neutral ISM
of I~Zw~18.

Summarizing, the chemical evolutionary history of I~Zw~18 suggests that the galaxy has experienced at least 
two star-formation events. Ancient activity (one or more star-formation episodes) is responsible for the metal 
enrichment of the H~{\sc i} gaseous component. The imprinting of the chemical elements produced by SNe~Ia is 
clearly detectable in this component, as inferred by the O/Fe ratio. A probable age of at 
least $\sim$ 1 Gyr can be inferred for this ancient star-formation event. More recent activity is instead 
accountable for the higher metal content in the H~{\sc ii} regions. It is more difficult to confine in time 
the extension of this stellar production. However, some constraints arise from the abundance patterns in the 
neutral and ionized gas. According to our findings, this event must be younger than $\sim$ 1 Gyr (no additional 
Fe enrichment in the H~{\sc ii} regions), but old enough to have enriched the gas in the central optical portion 
of the galaxy with primary nitrogen. This minimum age is 10 Myr or a few hundred Myr, depending on the source 
of primary nitrogen (massive or intermediate-mass stars). The scenario of a few hundred Myr is favored from 
considerations of time scales of mixing processes in the ISM.

\section{Summary and Conclusions}

We have presented our analysis of the interstellar spectrum of I~Zw~18, the most 
metal-poor star-forming galaxy in the local universe. The new $FUSE$ observations, 
with a spectral resolution of $\sim$ 35 km s$^{-1}$, have higher S/N and better 
coverage of the wavelength region 900$-$1200 \AA~ than previous observations. Our 
main findings are as follows.

1. We have studied the absorption lines of the H~{\sc i} Lyman series from Ly~$\beta$ to 
Ly~$\eta$. These lines are narrow in shape suggesting a purely interstellar origin, with 
no significant stellar contamination. A column density of 
$2.2^{+\,0.6}_{-\,0.5}\,\times\,10^{21}$ cm$^{-2}$ is inferred for H~{\sc i} in I~Zw~18 
by a multi-component fitting technique.

2. No H$_2$ lines are detected in I~Zw~18. We have set a conservative $3 \sigma$ upper limit 
of 5.25\,$\times$\,10$^{14}$ cm$^{-2}$ for the total column density of diffuse H$_2$. This
very low limit does not exclude the possible presence of clumpy H$_2$ in dusty compact 
star-forming regions that are opaque to far-UV radiation.

3. Many interstellar absorption lines of neutral and singly ionized atoms of
heavy elements have been detected and analyzed. Fe~{\sc ii} is the ion with the best 
column density constraint which is based on 9 lines with different oscillator strengths. 
Column density determinations of other ions, O~{\sc i}, Si~{\sc ii}, Ar~{\sc i}, and 
N~{\sc i}, have also been made. 

4. From the kinematics of the absorption lines, there is no clear evidence of gas 
infall/outflow over the large region sampled by $FUSE$. All the lines are centered 
at a velocity consistent with the systemic velocity of the galaxy as inferred by its 
stellar component ($v_{\rm stars}$\,=\,761 $\pm$ 9 km s$^{-1}$). 

5. The single- or two-velocity component models used in the multi-component fitting 
technique have turned out to be a quite good representation of the data when compared 
to the column densities inferred with the optical depth method. The ``effective'' Doppler 
width of the different ions is around 20  km s$^{-1}$, similar to our instrumental broadening.

6. We have inferred the following abundances of heavy elements: [Fe/H] = $-$\,1.76 $\pm$ 0.12, 
[O/H] = $-$\,2.06 $\pm$ 0.28, [Si/H] = $-$\,2.09 $\pm$ 0.12, [Ar/H] = $-$\,2.27 $\pm$ 0.13, and 
[N/H] = $-$\,2.88 $\pm$ 0.11. Ionization and dust depletion corrections have been considered and 
should be small. The neutral ISM in I~Zw~18 has thus been enriched in metals by a certain amount 
of star-formation.

7. A direct comparison of metal abundances in the neutral and ionized gas of I~Zw~18 indicates 
that the $\alpha$-elements (O, Ar, and Si) and N are several times lower in the neutral ISM, 
while the Fe content is the same within the errors. This suggests that the H~{\sc ii} regions 
have been enriched by additional star-formation.

8. The analysis of the abundance ratios O/Fe, N/O and O/Si and simple considerations of gas 
mixing, have allowed us to put some constraints on the chemical evolutionary history of I~Zw~18. 
It seems highly plausible that this galaxy has experienced two separate star-formation events. 
Ancient star-formation activity (one or more episodes) is responsible for the metal enrichment 
of the H~{\sc i} gas with the products typical of SNe~Ia. An age of at least $\sim$ 1 Gyr can 
thus be inferred. We estimated that the number of SNe~Ia required to produce the Fe content 
in the neutral ISM is $\sim$ 470. It is more difficult to confine in time the extension of the 
more recent star-formation activity responsible for the higher metal content in the H~{\sc ii} 
regions. If primary N is released by intermediate-mass (metal-poor massive) stars, then the 
enrichment requires several hundred Myr (10 Myr). The older age seems to be preferred by typical 
time scales related to mixing processes. In either case, the time scales are longer than the 
typical lifetime of H~{\sc ii} regions ($\lesssim$ 10 Myr).

It is worth noting that the simple scenario we inferred from the abundance patterns in 
the different gaseous components of I~Zw~18, is consistent with the star-formation history 
derived for this galaxy from the optical/NIR color-magnitude diagrams of its resolved stellar 
population (Aloisi et al. 1999; \"Ostlin 2000). These diagrams revealed for the first time 
the existence of single stars with ages around 1 Gyr, and the study presented in this paper 
seems to suggest the detection of their chemical imprinting in the neutral ISM of the galaxy. 

We can conclude that I~Zw~18 seems to be a dwarf star-forming galaxy which has substained a 
stellar-production activity for quite a long time, despite its very low metallicity. This 
activity has probably occurred with a low star-formation rate and/or relatively long quiescent 
periods. Hence, I~Zw~18 is just a relatively old galaxy with a very slow chemical evolution. 
This scenario is in agreement with the recent findings that all star-forming dwarf galaxies 
resolved into single stars show a stellar production that has gone on for quite a large 
fraction of a Hubble time with a so-called ``gasping'' star-formation regime (Tosi 2001). 

{\bf Acknowledgements}
The authors thank William van Dixon, Alex Fullerton, and Ravi Sankrit for their help in 
the calibration and spectral reduction of FUSE data. Fabrizio Brighenti, Thomas Brown, 
Annibale D'Ercole, Sally Oey, Livia Origlia, Nino Panagia, Simone Recchi, Monica Tosi, 
Giovanni Vladilo, Ren\'e Walterbos, and Rosemary Wyse are also acknowledged for stimulating 
discussion about I~Zw~18 and the chemical evolution of dwarf star-forming galaxies. We also 
thank Max Pettini, Evan Skillman, and Kim Venn for their careful reading of the manuscript
and their helpful suggestions for a more balanced paper. An anonymous referee is also 
acknowledged for his/her comments which contributed to improve the paper. This work is based 
on observations made with the NASA-CNES-CSA $FUSE$ mission operated for NASA by the Johns 
Hopkins University under NASA contract NAS5-32985. Financial support has been provided by 
NASA Long Term Space Astrophysics grants NAG5-6400 and NAG5-3485.

\clearpage

\clearpage

\clearpage 

\clearpage 

\clearpage 

\clearpage 

\clearpage 

\clearpage

\begin{deluxetable}{lccccc}
\tabletypesize{\footnotesize}
\tablewidth{0pc}
\tablenum{1}
\tablecaption{Parameters for H$_2$ intrinsic to I~Zw~18} 
\tablehead{
\colhead{Level} & \colhead{Line ID} & \colhead{$\lambda_{\rm lab}$} & 
\colhead{$\log$\,$(f\lambda)$~\tablenotemark{a}} & \colhead{$W_0$\,\tablenotemark{b} } & 
\colhead{$N_{\rm J}$\,\tablenotemark{c}} \\
\colhead{}       & \colhead{} & \colhead{(\AA)} & \colhead{} & \colhead{(m\AA)} & 
\colhead{(cm$^{-2}$)} }
\startdata
$J=0$ & (7$-$0) R(0) & 1012.817 & 1.483 & $<27.5$ &$<1.0\times10^{14}$\\
$J=1$ & (7$-$0) R(1) & 1013.441 & 1.307 & $<25.7$ &$<1.4\times10^{14}$\\
$J=2$ & (4$-$0) R(2) & 1051.498 & 1.168 & $<28.4$ &$<2.1\times10^{14}$\\
$J=3$ & (7$-$0) R(3) & 1017.427 & 1.263 & $<44.0$ &$<2.6\times10^{14}$\\
$J=4$ & (6$-$0) R(4) & 1032.351 & 1.244 & $<41.6$ &$<2.6\times10^{14}$\\
\enddata
\tablenotetext{a}{Wavelengths and oscillator strengths from Abgrall \& Roueff 
(1989).}
\tablenotetext{b}{Rest-frame equivalent width. Upper limits are 3\,$\sigma$.}
\tablenotetext{c}{Column density. Upper limits are 3\,$\sigma$.\\\\
N{\sc ote.}--- Lines in this table are those providing the most stringent limits 
to the inferred column densities in each $J$ level.}
\end{deluxetable}

\clearpage
		
\begin{deluxetable}{lllccclccccl}
\rotate
\footnotesize
\tablecaption{Interstellar Absorption Lines associated with I~Zw~18}
\tabletypesize{\scriptsize}
\tablewidth{0pt}
\tablenum{2}
\tablehead{
\colhead{} & 
\colhead{} & 
\colhead{} & 
\colhead{} & 
\multicolumn{3}{c}{P{\sc rofile} F{\sc itting}} & 
\multicolumn{1}{c}{\phm{gap}} & 
\multicolumn{2}{c}{A{\sc pparent} O{\sc ptical} D{\sc epth}} &
\multicolumn{1}{c}{\phm{gap}} & 
\colhead{} \\
\cline{5-7} \cline{9-10}
\colhead{Ion} &
\colhead{$\lambda_{\rm lab}$\,\tablenotemark{a}} &
\colhead{$f$\,\tablenotemark{b}} &
\colhead{Ref.} & 
\colhead{$z$} &
\colhead{$b$} &
\colhead{$\log$\,$N_{\rm PF}$} &
\colhead{} & 
\colhead{$\Delta v$} &
\colhead{$\log$\,$N_{\rm AOD}$} &
\colhead{} & 
\colhead{$W_0$} \\
\colhead{} &
\colhead{(\AA)} &
\colhead{} & 
\colhead{} & 
\colhead{} &
\colhead{(km s$^{-1}$)} &
\colhead{(cm$^{-2})$} &
\colhead{} &
\colhead{(km s$^{-1}$)} &
\colhead{(cm$^{-2})$} & 
\colhead{} &
\colhead{(m\AA)}
}
\startdata
Fe~{\sc ii}\,\tablenotemark{c} 
            & 1055.262  & 0.0075   & 2 & 0.0025462 & ~\,8.81 $\pm$ 0.47  & 15.00 $\pm$ 0.05 & & $-$\,40 to +\,80 & 14.91 $\pm$ 0.09 & &~\,56 $\pm$  9 \\
            &           &          &   & 0.0026983 & ~\,3.21 $\pm$ 0.38  & 14.37 $\pm$ 0.10 & &                  &                  & &              \\
            &           &          &   &   ...     &  ~\,...             & 15.09 $\pm$ 0.06\,\tablenotemark{d}
                                                                                            & &                  &                  & &              \\
            & 1063.176  & 0.05998  & 1 &           &                     &                  & & $-$\,40 to +\,80 & 14.21 $\pm$ 0.07 & &~\,80 $\pm$ 13 \\ 
            & 1112.048  & 0.0062   & 2 &           &                     &                  & & $-$\,40 to +\,80 & 15.12 $\pm$ 0.06 & &~\,78 $\pm$ 10 \\  
            & 1121.975\,\tablenotemark{e}
                        & 0.0202   & 2 &           &                     &                  & & $-$\,40 to +\,80 & 14.82 $\pm$ 0.04 & &  119 $\pm$ 11 \\
            & 1125.448  & 0.016    & 2 &           &                     &                  & & $-$\,40 to +\,80 & 14.85 $\pm$ 0.04 & &  110 $\pm$ 11 \\
            & 1133.665  & 0.0055   & 2 &           &                     &                  & & $-$\,40 to +\,80 & 15.23 $\pm$ 0.05 & &  100 $\pm$ 11 \\  
            & 1142.366  & 0.0042   & 2 &           &                     &                  & & $-$\,40 to +\,80 & 15.10 $\pm$ 0.09 & &~\,22 $\pm$  8 \\
            & 1143.226  & 0.0177   & 2 &           &                     &                  & & $-$\,40 to +\,80 & 14.92 $\pm$ 0.04 & &  145 $\pm$ 11 \\
            & 1144.938  & 0.106    & 2 &           &                     &                  & & $-$\,40 to +\,80 & 14.28 $\pm$ 0.03 & &  175 $\pm$ 11 \\
O~{\sc i}\,\tablenotemark{f}
            &~\,976.448\,\tablenotemark{g}
                        & 0.003300 & 1 & 0.0025786 & 23.00 $\pm$ 8.00  & 15.98 $\pm$ 0.26 & & $-$\,40 to +\,80 & 15.96 $\pm$ 0.12 & &  158 $\pm$ 31 \\
            & 1039.230  & 0.009197 & 1 &           &                     &                  & & $-$\,40 to +\,80 & 15.61 $\pm$ 0.05 & &  207 $\pm$ 13 \\
Si~{\sc ii} &~\,989.873\,\tablenotemark{h} 
                        & 0.1330   & 1 & 0.0025264 & 15.71 $\pm$ 2.85    & 14.81 $\pm$ 0.07 & &        ...       &        ...       & &  137 $\pm$ 17\,\tablenotemark{i} \\
            & 1020.699\,\tablenotemark{j}
                        & 0.02828  & 1 &           &                     &                  & & $-$\,40 to +\,30 & 14.70 $\pm$ 0.05\,\tablenotemark{k}
                                                                                                                                    & &~\,84 $\pm$  7\,\tablenotemark{i} \\
Ar~{\sc i}  & 1048.220\,\tablenotemark{l} 
                        & 0.263    & 3 & 0.0025246 &~\,27.02 $\pm$ 12.88 & 13.60 $\pm$ 0.08 & & $-$\,40 to +\,15 & 13.53 $\pm$ 0.05\,\tablenotemark{k}
                                                                                                                                    & &~\,78 $\pm$  8\,\tablenotemark{i} \\
            & 1066.660  & 0.0675   & 3 &           &                     &                  & & $-$\,40 to +\,80 & 13.64 $\pm$ 0.20 & &~\,29 $\pm$ 11 \\
N~{\sc i}   & 1134.415  & 0.02683  & 1 & 0.0025587 & 19.05 $\pm$ 4.83    & 14.44 $\pm$ 0.04 & & $-$\,40 to +\,80 & 14.51 $\pm$ 0.06 & &~\,69 $\pm$  9 \\
            & 1134.980  & 0.04023  & 1 &           &                     &                  & & $-$\,40 to +\,80 & 14.41 $\pm$ 0.05 & &  101 $\pm$ 10 \\
\enddata

\tablenotetext{a}{Vacuum wavelengths from Morton (1991).} 
\tablenotetext{b}{Oscillator strengths from references indicated in column 4.}
\tablenotetext{c}{Fit with two velocity components.}
\tablenotetext{d}{Total column density of the two velocity components.}
\tablenotetext{e}{Overlapping with HVC Fe~{\sc ii} $\lambda$1125, but contamination
                  negligible.}
\tablenotetext{f}{Line-profile fitting values of $N_{\rm PF}$ and $b$ for O~{\sc i} 
                  (and relative uncertainties) are to be intended as central value
                   of an interval of possible values (and relative half width). See
                   \S~7.1 for more details.}
\tablenotetext{g}{Continuum slightly affected by red wing of Ly~$\gamma$.}
\tablenotetext{h}{Possible blend with N~{\sc iii} $\lambda989$.}
\tablenotetext{i}{Equivalent width is a lower limit due to partial blend}
\tablenotetext{j}{Blend with H$_2$, continuum affected by blue wing of Ly~$\beta$.}
\tablenotetext{k}{Column density from the apparent optical depth method $N_{\rm AOD}$ is
                  a lower limit.}
\tablenotetext{l}{Blend with H$_2$.\\\\
R{\sc eferences.}---(1) Morton 1991. (2) Howk et al. 2000. (3) Morton 2003, in preparation.} 
\end{deluxetable}

\clearpage

\begin{deluxetable}{llcccccccc}
\rotate
\tabletypesize{\footnotesize}
\tablecaption{Interstellar Abundances in I~Zw~18}
\tablewidth{0pt}
\tablenum{3}
\tablehead{
\colhead{Element} &
\colhead{Ion} &
\colhead{} &
\colhead{$\log$$N$} &
\colhead{$\log$\,(X/H)} &
\colhead{$\log$\,(X/H)$_{\odot}$\,\tablenotemark{a}} &
\colhead{[X/H]\,\tablenotemark{b}} &
\colhead{} &
\multicolumn{2}{c}{[X/H]$_{\rm HII}$\,\tablenotemark{c}}\\
\cline{9-10}
\colhead{} &
\colhead{} &
\colhead{} &
\colhead{} &
\colhead{} &
\colhead{} &
\colhead{} &
\colhead{} &
\colhead{{\sc NW}} &
\colhead{{\sc SE}}
}
\startdata
H  & H~{\sc i}     & & 21.35 $\pm$ 0.10 &             ...      &    ...    &             ...      & &          ...         &           ...        \\
Fe & Fe~{\sc ii}   & & 15.09 $\pm$ 0.06\,\tablenotemark{d}
                                        & $-$\,6.26 $\pm$ 0.12 & $-$\,4.50 & $-$\,1.76 $\pm$ 0.12 & &          ...         & $-$\,1.96 $\pm$ 0.09 \\
O  & O~{\sc i}     & & 15.98 $\pm$ 0.26 & $-$\,5.37 $\pm$ 0.28 & $-$\,3.31 & $-$\,2.06 $\pm$ 0.28 & & $-$\,1.52 $\pm$ 0.03 & $-$\,1.51 $\pm$ 0.03 \\
Si & Si~{\sc ii}   & & 14.81 $\pm$ 0.07 & $-$\,6.54 $\pm$ 0.12 & $-$\,4.45 & $-$\,2.09 $\pm$ 0.12 & & $-$\,1.94 $\pm$ 0.23 & $-$\,1.86 $\pm$ 0.23 \\
Ar & Ar~{\sc i}    & & 13.60 $\pm$ 0.08 & $-$\,7.75 $\pm$ 0.13 & $-$\,5.48 & $-$\,2.27 $\pm$ 0.13 & & $-$\,1.43 $\pm$ 0.05 & $-$\,1.58 $\pm$ 0.05 \\
N  & N~{\sc i}     & & 14.44 $\pm$ 0.04 & $-$\,6.91 $\pm$ 0.11 & $-$\,4.03 & $-$\,2.88 $\pm$ 0.11 & &          ...         & $-$\,2.36 $\pm$ 0.07 \\
P  & P~{\sc ii}    & &    $<$ 13.60\,\tablenotemark{e}     
                                        &      $<$ $-$\,7.75   & $-$\,6.55 &      $<$ $-$\,1.20   & &          ...         &          ...         \\
\enddata
\tablenotetext{a}{Solar photospheric abundances from Grevesse et al. (1996), except for O which is from Allende Prieto 
                 et al. (2001).}
\tablenotetext{b}{[X/H] = $\log$\,(X/H) $-$ $\log$\,(X/H)$_{\odot}$.}
\tablenotetext{c}{Abundances in the northwest (NW) and southeast (SE) H~{\sc ii} regions from Izotov et al. (1999) and 
                  Izotov \& Thuan (1999). The same solar scale used for the neutral ISM has been adopted for consistency 
                  (column 5).} 
\tablenotetext{d}{Sum of the column densities of the two velocity components.}
\tablenotetext{e}{Upper limit is a $3 \sigma$ estimate inferred from P~{\sc ii} $\lambda$1152.}
\end{deluxetable}


\end{document}